\documentclass[journal,onecolumn,draft]{IEEEtran}
\IEEEoverridecommandlockouts

\usepackage{amsmath,bm}
\usepackage[final]{graphicx}
\usepackage{subfigure}
\usepackage{amsthm}
\usepackage{amssymb}
\usepackage{euscript}
\usepackage{setspace}
\usepackage{flushend,tikz,pgfplots,tkz-berge}

\newtheorem{theorem}{Theorem}
\newtheorem{lemma}{Lemma}

\theoremstyle{definition}
\newtheorem{definition}{Definition}
\newtheorem{example}{Example}

\newtheorem{remark}{Remark}

\newcommand{\defeq}{\overset{\text{def}}{=}}

\begin{document}

\title{On Cooperation in Multi-Terminal\\ Computation and Rate Distortion 
\thanks{This paper was presented in part at ISIT 2012.}
\thanks{This work was supported in part by a ``Future et Rupture'' grant from
the Institut Telecom, and by an Excellence Chair Grant
 from the French National Research Agency (ACE project).}
\thanks{ M.~Sefidgaran and A.~Tchamkerten are with the
Department of Communications and Electronics, Telecom
ParisTech, 46 Rue Barrault, 75634 Paris Cedex 13, France. Emails: \{sefidgaran,aslan.tchamkerten\}@telecom-paristech.fr.}}

\author
{
Milad Sefidgaran and Aslan Tchamkerten
}
\maketitle

\thispagestyle{empty}
\pagestyle{empty}

\begin{abstract} A receiver wants to compute a function of two
correlated sources separately observed by two  transmitters. One of the transmitters may send a possibly private message to the other transmitter in a cooperation phase before both transmitters
communicate to the receiver. For this network configuration this paper investigates both a function computation setup, wherein the receiver wants to compute a given function of the sources exactly,  and a rate distortion setup, wherein the receiver wants to compute a given function within some distortion. 

For the function computation setup, a general inner bound to the rate region is established and shown to be tight in a number of cases: partially
invertible functions, full cooperation between transmitters, one-round point-to-point
communication, two-round point-to-point communication, and the cascade setup where the transmitters and the receiver are aligned. In particular it is shown that the ratio of the total number of transmitted bits without cooperation and the total number of transmitted bits with cooperation can be arbitrarily large. Furthermore, one bit of cooperation suffices to arbitrarily reduce the amount of information both transmitters need to convey to the receiver.

For the rate distortion version, an inner bound to the rate region is exhibited which always  includes, and sometimes strictly, the convex hull of Kaspi-Berger's related inner bounds. The strict inclusion is shown via two examples. \end{abstract}

\section{ Introduction} 

Distributed function computation has been a long studied source coding problem in information theory. For identity functions, Shannon \cite{Shan49} and Slepian and Wolf \cite{SlepWol73} derived the rate regions for point-to-point and the noiseless multiple access configuration, respectively.  

Later, Gel'fand and Pinsker \cite{GelfPin79} considered the Slepian and Wolf's setting where an arbitrary function of the sources must be computed at the receiver. They proposed inner and outer bounds to the rate region which are tight in certain cases, including Shannon's point-to-point setting. However, the bounds are implicit and in general not computable.  In an independent and later work by Orlitsky and Roche \cite{OrliRoc01}, the rate region for the point-to-point setup is explicitly characterized using the concept of conditional characteristic graph defined and developed by K\"orner \cite{Korn73} and Witsenhausen \cite{Wits76}, respectively. Orlitsky and Roche also derived the rate region for two-round communication. This was recently generalized by Ma and Ishwar \cite{MaIsh11} to $m\geq 1$ communication rounds. 

Gel'fand and Pinsker's bounds are also tight for the multiple access configuration in the case where the sources are independent and in the case where the function is partially invertible.  These rate regions were recently made explicit using the concept of conditional characteristic graph in \cite{Sefi13} and \cite{SefiTchISIT11}, respectively.

At about the same time as Gel'fand and Pinsker's paper, K\"orner and Marton~\cite{KornMar79} considered the multiple access configuration where the receiver wants to compute the sum modulo two of two binary sources. They derived the rate region in the specific case where the sources have a symmetric distribution.  This result has later been generalized to sum modulo $p$, where $p$ is an arbitrary prime number \cite{HanKob87,ZhanVar98}. For sum modulo two and arbitrary distributions the best known inner bound is the one obtained Ahlswede and Han \cite{AhlsHan83}. Building on this work, Huang and Skoglund derived an achievable rate region for a certain class of polynomial functions which is larger than the Slepian-Wolf rate region \cite{HuanSko12,HuanSko12ISITA,HuanSko12CTW}. Finally, a variation of the problem where the receiver wants to compute some subspace generated by the sources has been investigated by Lalitha {\it{et al.}} \cite{LaliPraVin11}.

Another simple network configuration for which function computation has been investigated is the cascade network. The case with no side information at the receiver was investigated by Cuff {\it{et al.}} \cite{CuffSuElg09} and the case where the sources form a Markov chain was investigated by Viswanathan \cite{Visw10}. The general case was recently investigated in \cite{SefiTch122}.

 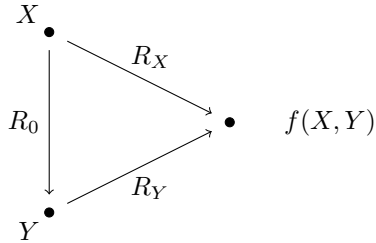
\begin{figure} 
  \centering
\begin{tikzpicture}[scale=1.2]
\draw [->] (0,1.8)--(0,.2);
\draw [->] (0.2,0.1)--(1.8,.9);
\draw [->] (0.2,1.9)--(1.8,1.1);
\draw [black,fill=black] (0,0) circle [radius=0.05];
\draw [black,fill=black] (0,2) circle [radius=0.05];
\draw [black,fill=black] (2,1) circle [radius=0.05];
\node [below left] at (0,0) {$Y$};
\node [above left] at (0,2) {$X$};
\node [right] at (2.5,1) {$f(X,Y)$};
\node [above right] at (.8,1.5) {$R_X$};
\node [left] at (0,1) {$R_0$};
\node [below right] at (.8,.5) {$R_Y$};
\end{tikzpicture}
  \caption{\label{fig:Problem}Function computation with cooperative transmitters.} 
\end{figure}
 
The aforementioned above mentioned point-to-point, multiple access, and cascade network configurations are all special cases of the network depicted in Fig.~\ref{fig:Problem}. Two sources $X$ and $Y$ are
separately observed by two transmitters, and a receiver wants to
compute  a function $f(X,Y)$ of the sources. Transmitter-$ X $
first sends some information to  transmitter-$ Y $ at rate $R_0$ (cooperation or private
phase), then transmitter-$ X $ and transmitter-$ Y $ send  information to the receiver at rate $R_X$ and $R_Y$, respectively.\footnote{The case where the sent message from transmitter-$X$ to transmitter-$Y$ can be overheard by the receiver was considered in a rate distortion setup by Kaspi and Berger \cite{KaspBerg82} and in a function computation setup by Ericsson and K\"orner \cite{EricKor83}.}
This paper investigates this setting in the context of both function computation and rate distortion.

The first part of the paper is devoted to function computation. The main result is a general  inner bound that is tight in a number of special cases: 
\begin{itemize}
\item 
the function is partially invertible---{\it{i.e.}}, when $X$ is a function of $f(X,Y)$;
\item 
unlimited cooperation, {\it{i.e.}}, when transmitter-$ Y $ knows $ X $, for which we recover the results of \cite{AhlsKor75}; 
\item
one and two-round point-to-point communication for which we recover the results in \cite{OrliRoc01};
\item
cascade network for which we recover the results of
\cite{CuffSuElg09,Visw11};
\item
no cooperation: invertible function or independent sources for which we recover the results of \cite{SefiTchISIT11}.
\end{itemize}
We also establish through an example that one bit of cooperation can arbitrarily reduce the amount of information both transmitters need to convey to the receiver. Specifically, the ratio
$$\frac{\min \limits_{\text{no cooperation}} \quad R_X+R_Y}{\min \limits_{\text{with cooperation}} \quad R_X+R_Y+R_0}$$
 can be made arbitrarily large. This is analogous to the Orlitsky's results \cite{Orli90,Orli91} for interactive point-to-point communication that says that the number of bits needed for one-round communication can be arbitrarily larger than the number of bits needed for two-round communication. 
 
In the second part of the paper we consider the problem where the receiver wants to recover some functions $f_1(X,Y)$ and $f_2(X,Y)$ within some distortions. 
For the special case where $f_1(X,Y)=X$ and $f_2(X,Y)=Y$ Kaspi and Berger \cite{KaspBerg82} proposed two inner bounds. The first is a general inner bound while the second holds and is tight in the full cooperation case only. These bounds easily generalize to arbitrary functions by using similar arguments as those used by Yamamoto in \cite[Proof of Theorem~1]{Yama82}
 to extend Wyner and Ziv's result \cite[Theorem~1]{WyneZiv76} from identity functions to arbitrary functions.

Finally, building on achievability arguments used to establish the inner bound for the function computation problem we
derive a new inner bound for the corresponding rate distortion
setup. This inner bound always includes, and in certain cases strictly, the convex hull of 
Kaspi-Berger's inner bounds \cite[Theorems 5.1 and 5.4]{KaspBerg82} generalized to arbitrary functions. Two examples are given to illustrate strict inclusion.

The paper is organized as follows. Section
\ref{sec:ProbState} contains the problem statement and
provides some background material and definitions. Section \ref{sec:mainResults} contains the results and
Section~\ref{sec:analysis} is devoted to their proofs.

\section{Problem Statement and Preliminaries} \label{sec:ProbState} 
We use calligraphic fonts to denote the range of the corresponding random variable. For instance, $\cal{X}$ denotes the range of $X$ and $\cal{T}$ denotes the range of $T$. 

Let $\mathcal{X}$, $\mathcal{Y}$, $\mathcal{F}$, $\mathcal{F}_1$, and $\mathcal{F}_2$ be finite sets. Further, define $$f:\mathcal{X}\times\mathcal{Y} \rightarrow \mathcal{F}$$ $$f_i:\mathcal{X}\times\mathcal{Y} \rightarrow \mathcal{F}_i\quad i\in \{1,2\}$$ $$d_i:\mathcal{F}_i\times \mathcal{F}_i \rightarrow \mathbb{R}^+ \quad i\in \{1,2\} \,.$$ 

Let $\{ (x_i,y_i)\}_{i=1}^ {\infty}$ be independent instances of
random variables $(X,Y)$ taking values
over $ \mathcal{X} \times \mathcal{Y} $ and
distributed according to $p(x,y)$. Define $$\mathbf{X}\defeq X_1,\ldots,X_n$$ and 
$$f(\mathbf{X},\mathbf{Y})\defeq f(X_1,Y_1),...,f(X_n,Y_n) \,.$$
Similarly define $f_1(\mathbf{X},\mathbf{Y})$ and $f_2(\mathbf{X},\mathbf{Y})$.

Next, we recall the notions of achievable rate tuples for function computation and rate distortion. For function computation it is custom to consider asymptotic zero block error probability whereas for rate distortion it is custom to consider bit average distortion.

\begin{definition}[Code]
An $(n,R_0,R_X,R_Y)$ code for the function computation setup where the receiver wants to compute $f$ consists of three encoding functions
 \begin{align*}  \varphi_0&:\mathcal{X}^n \rightarrow \{ 1,2,..,2^{nR_0} \}
\nonumber \\ \varphi_X&:\mathcal{X}^n \rightarrow \{ 1,2,..,2^{nR_X} \} 
\nonumber \\ \varphi_Y&:\mathcal{Y}^n \times \{ 1,2,..,2^{nR_0} \} \rightarrow \{ 1,2,..,2^{nR_Y} \}
\end{align*} 
and a decoding function
 \begin{align*}  
\psi&: \{ 1,2,..,2^{nR_X} \} \times  \{ 1,2,..,2^{nR_Y} \} 
\rightarrow \mathcal{F}^n\,. \nonumber 
\end{align*} 
The corresponding error probability is defined as
$$P(\psi(\varphi_X(\mathbf{X}),\varphi_Y(\varphi_0(\mathbf{X}),\mathbf{Y}))\ne f(\mathbf{X},\mathbf{Y})).$$

An $(n,R_0,R_X,R_Y)$ code for the rate distortion problem where the receiver wants to compute $f_1$ and $f_2$ consists of three encoding functions defined as for the function computation problem, and two decoding functions
 \begin{align*}  
\psi_i&: \{ 1,2,..,2^{nR_X} \} \times  \{ 1,2,..,2^{nR_Y} \} 
\rightarrow \mathcal{F}_i^n\quad i \in \{1,2\}\,. \nonumber \end{align*} 
The corresponding average distortions are defined as\footnote{We use $\mathbb{E}$ to denote expectation.} 
$$\mathbb{E}d_i(f_i(\mathbf{X},\mathbf{Y}),\psi_i(\varphi_X(\mathbf{X}),\varphi_Y(\varphi_0(\mathbf{X}),\mathbf{Y})))=\frac{1}{n}\sum \limits_{j=1}^n \mathbb{E}d_i(f_i(X_j,Y_j),\psi_i(\varphi_X(\mathbf{X}),\varphi_Y(\varphi_0(\mathbf{X}),\mathbf{Y}))_j)$$
for $i \in \{1,2\}$. In the above expression $\psi_i(\cdot,\cdot)_j$ refers to the $j$\/th component of the length $n$ vector $\psi_i(\cdot,\cdot)$.
\end{definition}
\begin{definition}[Function Computation Rate Region]\label{rateComputationDef} 
A rate tuple $(R_0,R_X,R_Y)$  is said to be achievable if, for any $\varepsilon > 0$ and all $n$
large enough, there exists an $(n,R_0,R_X,R_Y)$ code whose error probability is no
larger than~$\varepsilon$. The rate region is the closure of the set of achievable 
rate tuples $(R_0,R_X,R_Y)$.
\end{definition}
\begin{definition}[Rate Distortion Rate Region]\label{rateDistortionDef}
Let $D_1$ and $D_2$ be two non-negative constants. A rate tuple $(R_0,R_X,R_Y)$  is said to be achievable with distortions $D_1$ and $D_2$ if, for any $\varepsilon > 0$ and all $n$
large enough, there exists an $(n,R_0,R_X,R_Y)$ code whose average distortions are no larger than $D_1$ and $D_2$, respectively. The rate distortion rate region with respect to $D_1$ and $D_2$ is the closure of the set of achievable rate tuples $(R_0,R_X,R_Y)$ with distortions $D_1$ and $D_2$.
\end{definition}

The problems we consider are the characterizations of
\begin{itemize}
\item[i. ] the function computation rate region for given function$f(x,y)$ and distribution~$p(x,y)$; 
\item[ii. ]  the rate distortion rate region for given functions $f_1(x,y)$, $f_2(x,y)$, distribution $p(x,y)$, and distortion constraints  $D_1$, $D_2$. 
\end{itemize}

Conditional characteristic graphs play a key role in function computation problem \cite{Wits76,Korn73,SefiTchISIT11}. Below we introduce a general definition of 
conditional characteristic graph.
 
\begin{remark} Given two random variables $X$ and $V$, where $X$ ranges over
$\cal{X}$ and $V$ over {\emph{subsets}} of $\cal{X}$,\footnote{{\it{I.e.}}, a sample
of $V$ is a subset of $\cal{X}$.} we write $ X \in V $
whenever $P(X\in V)=1$.
\end{remark}

Recall that an independent set of a graph $G$ is a subset of vertices no two of
which are connected. The set of independent sets of $G$ is 
denoted by $\Gamma(G)$.
\begin{definition}[Generalized Conditional Characteristic Graph]
\label{def:CondCharGraph} Let $L$, $K$, and $S$ be arbitrary
discrete random variables with $(L,K,S)\sim p(l,k,s)$. Let $f:\mathcal{S}\rightarrow \mathbb{R}$ be a function such that $H(f(S)|L,K)=0$. The conditional characteristic
graph $G_{L|K}(f)$ of $L$ given $K$ with respect to the function $f(s)$ is the graph whose vertex set is
$\mathcal{L} $ and such that $l_1 \in \mathcal{L}$ and $l_2 \in \mathcal{L}$ are connected if for some $s_1,s_2 \in \mathcal{S}$,  and $k \in \mathcal{K}$
 \begin{itemize} 
\item[i.] $p(l_1,k,s_1) \cdot p(l_2,k,s_2) > 0$, 
\item[ii.] $f(s_1)\neq f(s_2)$. 
\end{itemize}
When there is no ambiguity on the function $f(s)$, the above conditional characteristic graph is simply denoted by $G_{L|K}$.
\end{definition}

\begin{definition} [Conditional Graph Entropy \cite{OrliRoc01}]
\label{def:condgraph} Given $(L,K,S)\sim p(l,k,s)$ and $f:\mathcal{S}\rightarrow \mathbb{R}$  such that $H(f(S)|L,K)=0$, the conditional graph entropy $H(G_{L|K}(f))$ is defined as 
 \begin{align*}
H(G_{L|K}(f))\defeq \min \limits_{\substack{ V -
L- K\\ L \in V \in \Gamma(G_{L|K})}} I(V;L|K)
\end{align*}
where $V-L-K$ refers to the standard Markov chain notation.
\end{definition}

\section{Results} \label{sec:mainResults}
In the first part of this section we consider the function computation problem formulation and in the second part of the section we consider the corresponding rate distortion formulation.
\subsection{Computation}

Our first result is a general inner bound to the function computation rate
region (see Definition~\ref{rateComputationDef}).

\begin{theorem}[Inner Bound -- Computation] \label{achiev} $ (R_0,R_X,R_Y) $ is achievable whenever 
\begin{align} R_0 &> I(X;U|Y) \nonumber \\ 
R_X &> I(V;X|T,W) \nonumber \\  
R_Y &> I(U,Y;W|V,T) \nonumber \\  
R_X+R_Y &> I(X,Y;V,T,W)+I(U;W|V,X,T,Y),\label{eq:th1Rates}
\end{align} for some $ T $, $ U $, $V$, and $W$ that satisfy
\begin{align}
&T-U-X-Y \nonumber \\
V - (&X,T) - (U,Y)-W, \label{eq:th1Markovs}
\end{align}
and  
\begin{align}
 (T,X)&\in V \in \textnormal{M}(\Gamma(G_{T,X|T,U,Y})) \nonumber \\
(T,U,Y&) \in W \in \textnormal{M}(\Gamma(G_{T,U,Y|T,V}))\,. \label{eq:th1Sets}
\end{align}
Moreover, the following cardinality bounds hold  \begin{align}
|\mathcal{T}| & \leq |\mathcal{X}|+4 \nonumber \\
|\mathcal{V}| & \leq (|\mathcal{X}|+4)\cdot |\mathcal{X}|+1 \nonumber \\
|\mathcal{W}| & \leq |\mathcal{U}| \cdot |\mathcal{Y}|+1. \label{eq:th1Cardinalities}
\end{align}
on the alphabets $\mathcal{T}$, $\mathcal{U}$, $\mathcal{V}$, $\mathcal{W}$ of $ T $, $ U $, $V$, $W$, respectively.
\end{theorem}
The last part of the theorem says that the achievable rate region \eqref{eq:th1Rates} is maximal for random  variables $T$, $V$, and $W$ defined over sets whose cardinalities are bounded as in \eqref{eq:th1Cardinalities}. Note that in the graphs $G_{T,X|T,U,Y}$ and $G_{T,U,Y|T,V}$ the random variable $T$ can be interpreted as a common randomness over a set of conditional characteristic graphs.

The rate region characterized in Theorem~\ref{achiev} turns out to be tight in a number of
interesting cases which we now list. The first case holds when the function is partially
invertible with respect to $X$, {\it{i.e.}}, when $X$
is a function of $f(X,Y)$.
\begin{theorem}[Partially Invertible Function] \label{rate:partially}
The inner bound is tight when
$f(X,Y) $ is partially invertible with respect to $ X $. In this case, the rate
region reduces to 
\begin{align} R_0 &\geq I(X;U|Y) \nonumber \\ 
 R_X &\geq H(X|U,W) \nonumber \\ R_Y &\geq
I(Y;W|X,U) \nonumber \\ R_X
+R_Y &\geq H(X)+I(Y;W|U),\nonumber 
\end{align} for some $ U $ and $ W $ with alphabets $\mathcal{U}$ and $\mathcal{W}$, respectively, that satisfy
\begin{align}
U&-X-Y \nonumber \\
X - &(U,Y)-W, \label{eq:rateCPartiallyMarkovs}
\end{align}
and  
\begin{align}
(U,Y) \in W \in \textnormal{M}(\Gamma(G_{U,Y|X,U}))\,. \label{eq:rateCPartiallySets}
\end{align}
Moreover, the following cardinality bounds hold
\begin{align}
|\mathcal{U}| & \leq |\mathcal{X}|+4 \nonumber \\
|\mathcal{W}| & \leq (|\mathcal{X}|+4) \cdot |\mathcal{Y}|+1.  \label{eq:rateCPartiallyCardinality}
\end{align}
\end{theorem}

In the following example we apply Theorem~\ref{rate:partially} to show that 
\begin{itemize}
\item one bit of cooperation can
arbitrarily reduce the amount of information both transmitters need to convey to the receiver,
\item the ratio of the minimum sum rate without cooperation and the minimum sum rate with cooperation can be arbitrary large.
\end{itemize}

\begin{example}\label{example:sumratevscoo}
\begin{figure}  \centerline{\includegraphics[scale=0.3]{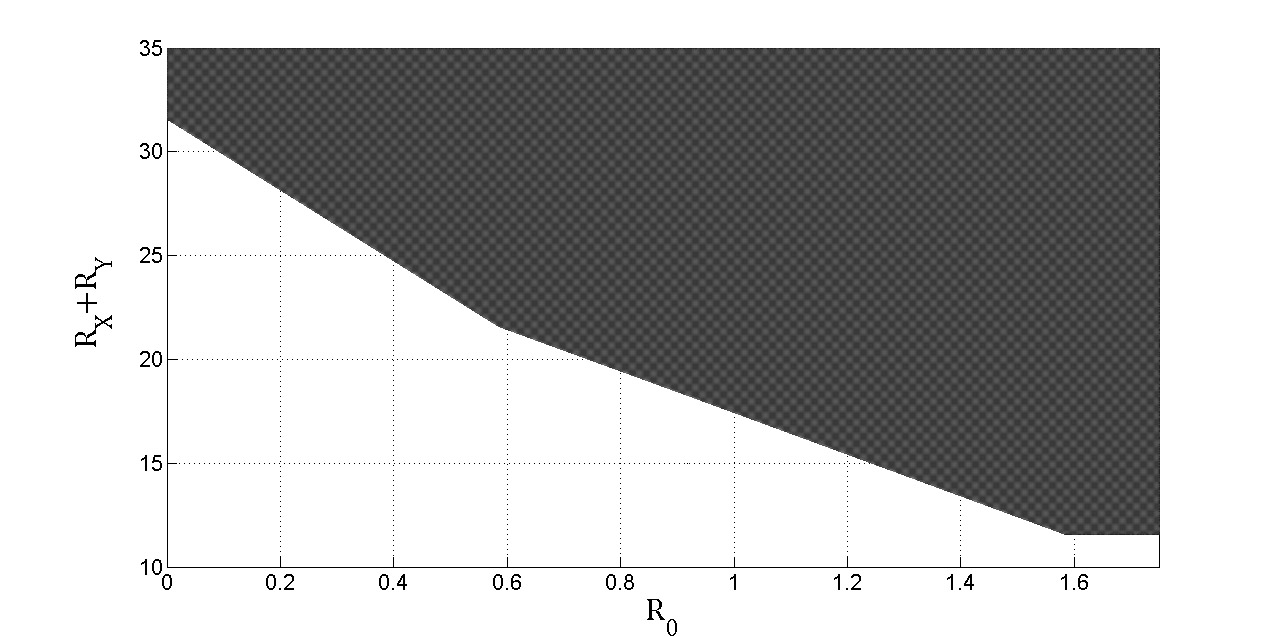}} \caption{\label{fig:CooperativepartiallyInv}
Minimum sum-rate $R_X+R_Y$ as a function of the cooperation rate $R_0$ for the partially invertible function of Example~\ref{example:sumratevscoo} with $a=3$ and $b=10$. } \end{figure}
 
Let $a\geq 2$ and $b\geq 1$ be two natural numbers. Let $X$ be uniform over $\{1,2,\cdots,a\}$ and let $Y=(Y_1,Y_2,\cdots,Y_a)$ where the $Y_{i}'s, i \in \{1,2,\cdots,a\}$, are independent random variables, each of them uniformly distributed over $\{1,\cdots,2^b\}$ and independent of $X$. The receiver wants to recover $X$ and $Y_X$, \textit{i.e.}, $f(X,Y)=(X,Y_X)$.

From Theorem \ref{rate:partially} and the fact that $X$ and $Y$ are independent the rate is given by\begin{align*}
R_0 &\geq I(X;U) \\
R_X & \geq H(X|U) \\
R_Y & \geq I(Y;W|U) \\
R_X+R_Y & \geq H(X)+I(Y;W|U) 
\end{align*}
for some $U$ and $W$ that satisfy \eqref{eq:rateCPartiallyMarkovs}, \eqref{eq:rateCPartiallySets}, and \eqref{eq:rateCPartiallyCardinality}.

We evaluate the sum rate constraint. Since $X$ is uniformly distributed we have $H(X)=\log \limits_2(a)$. Now, due to the independence of $X$ and $Y$ and the Markov chain $U-X-Y$ we have 
\begin{align}
H(Y|U)=H(Y)=a\cdot b\,. \label{eq:expartially2}
\end{align}

Further, by Definition~\ref{def:CondCharGraph}, for each $u \in \mathcal{U}$, $(u,y)=(u,(y_1,y_2,\cdots,y_a))$ and $(u,y')=(u,(y'_1,y'_2,\cdots,y'_a))$ are connected in $G_{U,Y|X,U}$ if and only if $y_x \neq y'_x$ for some 
$$x\in {\cal{A}}_u \defeq \{x:p(u,x)>0\}.$$
Hence, because $W$ satisfies \eqref{eq:rateCPartiallySets}, conditioned on $U=u$ the maximum number of elements in an independent set $w \in W$ that contains vertices $(u,y)$, $y \in \mathcal{Y}$, is $ 2^{b(a-|{\cal{A}}_u|)}$.\footnote{We use $|{\cal{A}}_u|$ to denote the cardinality of ${\mathcal{A}}_u$.} Therefore,
\begin{align}
H(Y|W,U=u)= b \cdot  (a-|{\cal{A}}_u|) \label{eq:expartially1}
\end{align}
by letting $W$ take as values maximal independent sets. 

Equations \eqref{eq:expartially2} and \eqref{eq:expartially1} give
 $$\min \limits_{W} I(Y;W|U)=b \cdot \sum \limits_{u \in \mathcal{U}} |{\cal{A}}_u| \cdot p(u)\,, $$
 and therefore
 \begin{align}
R_0 &= \log \limits_2(a) + \sum \limits_{x,u} p(x,u)\cdot \log \limits_2 p(x|u)\nonumber \\
R_X+R_Y & = \log \limits_2(a)+b \cdot \sum \limits_{u \in \mathcal{U}} |{\cal{A}}_u| \cdot p(u)\label{rosumrate}
\end{align}
for any valid choice of $U$. 
By considering all random variables $U$ over alphabets of no more than $a+4$ elements and that satisfy the Markov chain $U-X-Y$, one can numerically evaluate the minimum achievable sum rate for all values of $R_0$ using the above equations. Fig.~\ref{fig:CooperativepartiallyInv} shows the minimum achievable sum rate $R_X+R_Y$ as a function of $R_0$  for $a=4$ and $b=10$.

Choosing $U \in \{0,1\}$ in \eqref{rosumrate} such that 
\begin{align*}
p(U=0|X=1)=p(U=0|X=2)=p(U=1|X=3)=p(U=1|X=4)=1 \\
p(U=0|X=3)=p(U=0|X=4)=p(U=1|X=1)=p(U=1|X=2)=0
\end{align*}
shows that
\begin{align}
R_0 &=1 \nonumber \\
R_X+R_Y & = 2+  2 \cdot b \label{1}
\end{align}
is achievable. 

When $R_0=0$ the minimum sum rate is given by
\begin{align}
\min(R_X+R_Y )& = 2+  4 \cdot b\label{2}
\end{align}
from \cite{SefiTchISIT11} and using the fact that the function is partially invertible and that the sources are independent.\footnote{More generally, one can easily check that the minimum sum rate without cooperation is $\log \limits_2(a)+a \cdot b $.}

From \eqref{1} and \eqref{2} we deduce that one bit of cooperation decreases the sum rate by at least $2 \cdot b$, which can be arbitrarily large since $b$ is an arbitrary natural number.

Moreover with $R_0=\log \limits_2(a)$ bits of cooperation, one can see that by letting $U=X$,
$$\min \limits_{\text{with cooperation}} \quad R_X+R_Y+R_0 \leq 2\log\limits_2(a)+b.$$
Hence,
$$\frac{\min \limits_{\text{no cooperation}} \quad R_X+R_Y}{\min \limits_{\text{with cooperation}} \quad R_X+R_Y+R_0} \geq \frac{\log\limits_2(a)+a\cdot b}{2\log\limits_2(a)+b},$$
which can be arbitrary large.
 \end{example}

The next theorems provide three other cases where Theorem~\ref{achiev} is tight. For each of them one of the links is rate unlimited. 

\begin{figure} 
\centering
\subfigure[]{
\begin{tikzpicture}[scale=1]
\draw [->] (0.2,0.1)--(1.8,.9);
\draw [->] (0.2,1.9)--(1.8,1.1);
\draw [black,fill=black] (0,0) circle [radius=0.05];
\draw [black,fill=black] (0,2) circle [radius=0.05];
\draw [black,fill=black] (2,1) circle [radius=0.05];
\node [below left] at (0,0) {$(X,Y)$};
\node [above left] at (0,2) {$X$};
\node [right] at (2.2,1) {$f(X,Y)$};
\node [above right] at (.8,1.5) {$R_X$};
\node [below right] at (.8,.5) {$R_Y$};
\end{tikzpicture}
\label{fig:SWR0}
}
\hspace{3 cm}
\subfigure[]{
\begin{tikzpicture}[scale=1]
\draw [->] (0.2,0.1)--(1.8,0.1);
\draw [->] (1.8,-0.1)--(.2,-0.1);
\draw [black,fill=black] (0,0) circle [radius=0.05];
\draw [black,fill=black] (2,0) circle [radius=0.05];
\node [left] at (-0.05,0) {$Y$};
\node [right] at (2.05,0) {$X$};
\node [above] at (1,0.1) {$R_Y$};
\node [below] at (1,-0.1) {$R_0$};
\node [right] at (2.8,0) {$f(X,Y)$};
\node  at (0,-1.5) {};
\end{tikzpicture}
\label{fig:SWX}
}
\\
\subfigure[]{
\begin{tikzpicture}[scale=1]
\draw [->] (0.2,0.1)--(1.8,0.1);
\draw [->] (.2,-0.1)--(1.8,-0.1);
\draw [black,fill=black] (0,0) circle [radius=0.05];
\draw [black,fill=black] (2,0) circle [radius=0.05];
\node [left] at (-0.05,0) {$X$};
\node [right] at (2.05,0) {$Y$};
\node [above] at (1,0.1) {$R_X$};
\node [below] at (1,-0.1) {$R_0$};
\node [right] at (2.8,0) {$f(X,Y)$};
\end{tikzpicture}
\label{fig:SWY}
}
\hspace{2 cm}
\subfigure[]{
\begin{tikzpicture}[scale=1]
\draw [->] (0.2,0)--(1.8,0);
\draw [->] (2.2,0)--(3.8,0);
\draw [black,fill=black] (0,0) circle [radius=0.05];
\draw [black,fill=black] (2,0) circle [radius=0.05];
\draw [black,fill=black] (4,0) circle [radius=0.05];
\node [below] at (0,0) {$X$};
\node [below] at (2,0) {$Y$};
\node [right] at (4.8,0) {$f(X,Y)$};
\node [above] at (1,0) {$R_0$};
\node [above] at (3,0) {$R_Y$};
\node  at (0,-.2) {};
\end{tikzpicture}
\label{fig:Cascade}
}

\label{fig:subSettings}
\caption{
\subref{fig:SWR0} Full cooperation,
\subref{fig:SWX} Two round point-to-point communication, 
\subref{fig:SWY} One round point-to-point communication,
\subref{fig:Cascade} Cascade.}
\end{figure}
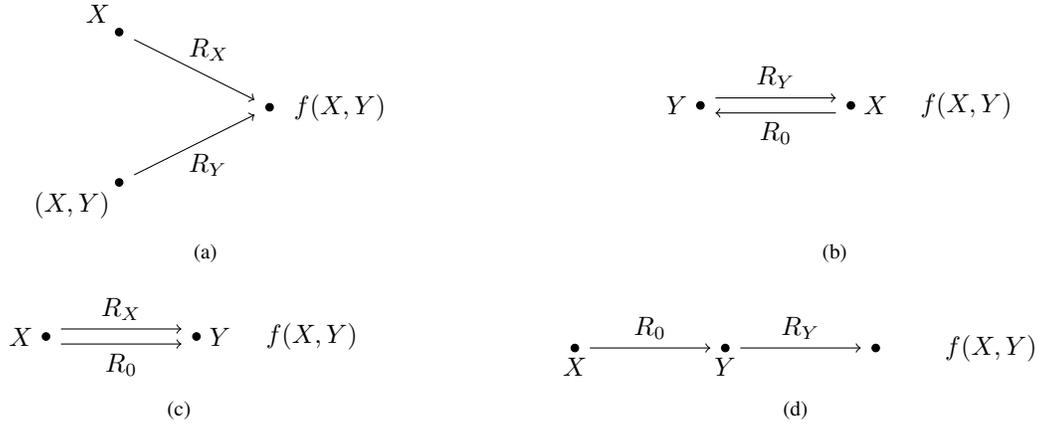

When there is full cooperation
between transmitters,
{\it{i.e.}}, when transmitter-$ Y $ has full access to source
$X$, the setting is captured
by the condition $R_0> H(X|Y)$ and is depicted in  Fig. \ref{fig:SWR0}.  In this case, we recover the results of \cite[Theorem~2]{AhlsKor75} and \cite[Theorem~2.1]{Wyne75} with improved cardinality bound.
\begin{theorem}[Full Cooperation] \label{rate:R0}
The inner bound is tight when
$$ R_0 > H(X|Y).$$ In this case, the rate region reduces
to 
\begin{align} R_0 &\geq H(X|Y) \nonumber \\ 
  R_Y &\geq H(f(X,Y)|T) \nonumber \\  
  R_X+R_Y &\geq H(f(X,Y))+I(X;T|f(X,Y)),\nonumber 
\end{align} 
for some $ T $ with alphabet $\mathcal{T}$ that satisfies
$$T-X-Y,$$
with cardinality bound 
\begin{align*}
|\mathcal{T}| & \leq |\mathcal{X}|+1\,.
\end{align*}

\end{theorem}

In the following example, we derive the rate region for a partially invertible function when there is no cooperation and when there is full cooperation. 
\begin{example}
\begin{figure}  \centerline{\includegraphics[scale=0.3]{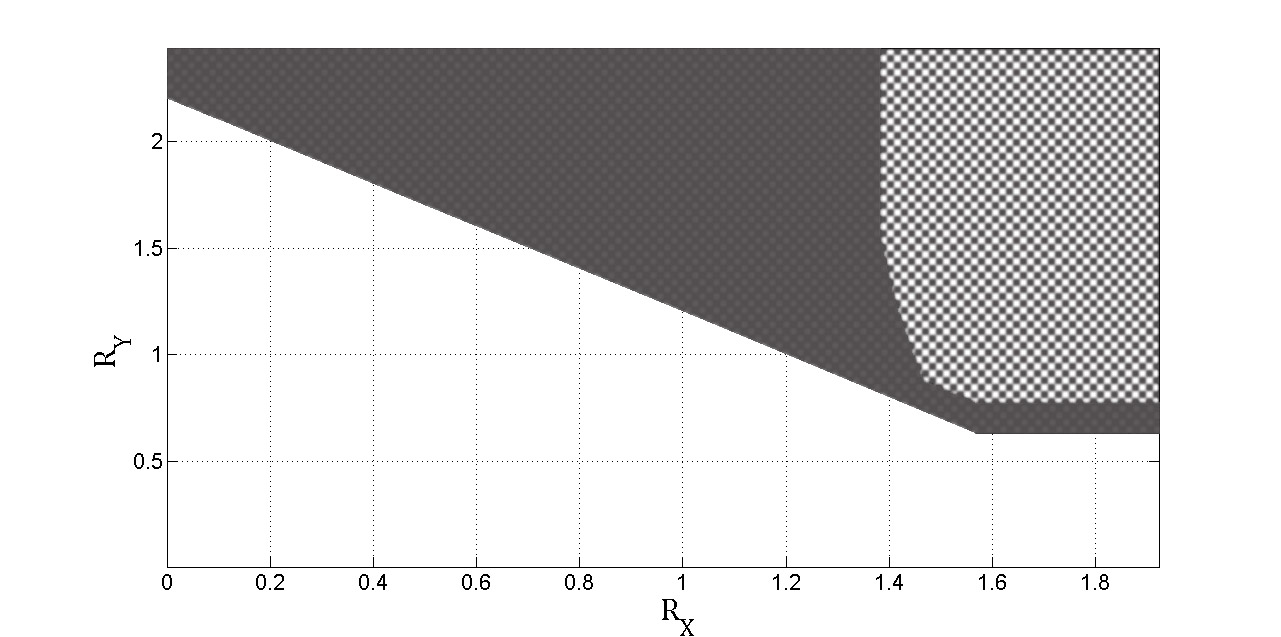}} \caption{\label{fig:rateFullCooperation}
Example of the rate region for a partially invertible function for $R_0=0$ and $R_0=H(X|Y)$.} \end{figure}
Let $f(x,y)= (-1)^y \cdot x $,
with $\mathcal{X}=\mathcal{Y}=\{0,1,2\}$,  and
$$p(x,y)= \left[ \begin{array}{ccc} .21& .03& .12 \\ .06& .15& .16 \\  .03& .12& .12 \end{array}
\right].$$  The rate region when $R_0=0$ can be derived from an example in \cite{SefiTchISIT11} which considers the function $g(x,y)=(y \text{ mod } 2)+3x$. Since there is a one-to-one mapping between this function and $f(x,y)$, this latter can be computed if and only if $g(x,y)$ can be computed. Hence these two functions have the same rate region which is depicted by the gray area in Fig. \ref{fig:rateFullCooperation}. 

Under full cooperation, \textit{i.e.}, $R_0=H(X|Y)=1.38$,  Theorem \ref{rate:R0} provides the rate region which is given by the union of the gray and the black areas in Fig. \ref{fig:rateFullCooperation}. Note that the black area, which represents the difference between the two regions, is non-symmetric with respect to $X$ and $Y$, as can be expected.
\end{example}

When $R_Y$ is
unlimited, {\it{i.e.}}, when the receiver looks over the
shoulders of transmitter-$Y$, the setting is captured
by condition $R_Y> R_0+H(Y)$ and reduces to point-to-point
communication. This is depicted in Fig.~\ref{fig:SWY} with the transmitter
observing $X$ and the
receiver observing $Y$. The rate region for this case was  
established in \cite[Theorem~1]{OrliRoc01}.
\begin{theorem}[One-Round Point-to-Point Communication]\label{rate:RY}
The inner bound is tight when
$$R_Y > R_0 + H(Y)\,.$$ In this case, the rate region reduces to
$$R_0+R_X\geq H(G_{X|Y})\,.$$\end{theorem}

When condition $R_X> H(X)$ holds, the situation reduces to the two-round communication setting
depicted in Fig.~\ref{fig:SWX}. The receiver, having access to $X$, first conveys information to
transmitter-$ Y $ which then replies.
\begin{theorem}[Two-Round Point-to-Point Communication] \label{rate:RX}
The inner bound is tight when
 $$ R_X > H(X)\,. $$
 In this case, the rate region reduces to
\begin{align} R_0 &\geq I(X;U|Y) \nonumber \\ 
R_X &\geq H(X) \nonumber \\
R_Y &\geq I(Y;W|X,U) \label{eq:twoRoundRates}
\end{align} for some $ U $ and $W$ with alphabets $\mathcal{U}$ and $\mathcal{W}$, respectively, that satisfy
\begin{align*}
&U-X-Y \\
X&-(U,Y)-W,
\end{align*}
and
$$(U,Y) \in W \in \textnormal{M}(\Gamma(G_{U,Y|X,U})),$$ 
with cardinality bounds 
\begin{align*}
|\mathcal{U}| & \leq |\mathcal{X}|+2 \\
|\mathcal{W}| & \leq (|\mathcal{X}|+2)\cdot |\mathcal{Y}|+1 .
\end{align*}

 \end{theorem}
The rate region in the above case $R_X > H(X)$ was previously established in
\cite[Theorem~3]{OrliRoc01}, and has been generalized in \cite[Theorem~1]{MaIsh11} for $m$-round point-to-point communication with cardinality bounds on the alphabet of auxiliary variables. However in both works, the range of the auxiliary random
variable $W$ was left unspecified, except for the condition that $U,W,X$ should determine
$f(X,Y)$. By contrast, Theorem~\ref{rate:RX} specifies
$W$ to range over independent sets of a suitable graph. Also, the cardinality bounds are tighter with respect to the bounds derived in~\cite[Theorem~1]{MaIsh11}.

Finally, when $R_X=0$ there is no direct link between transmitter-$ X $
and the receiver and the situation reduces to the cascade setting depicted in
Fig.~\ref{fig:Cascade}. The rate region for this case was  
established in \cite[Theorem 3.1]{CuffSuElg09} (see also \cite[Theorem 2]{Visw11}).\footnote{For the corresponding rate distortion problem, look at \cite{CuffSuElg09} and \cite{PermWei12}.}
\begin{theorem}[Cascade]\label{rate:Cascade}
The inner bound is tight when $$ R_X=0. $$ In this case, the rate region reduces
to \begin{align*} R_0&\geq H(G_{X|Y})\\
R_Y&\geq H(f(X,Y))\,.
\end{align*}
\end{theorem}

\subsection{Rate Distortion}   Theorem~\ref{achiev}
gives an inner bound to the rate distortion problem (see Definition \ref{rateDistortionDef})
with zero distortions when both distortion functions are the same. It
turns out that this inner bound is in general larger
than the rate region obtained by Kaspi and Berger in
\cite[Theorem 5.1]{KaspBerg82} for zero distortions. 
The reason for this lies in Kaspi and Berger's achievable
scheme which their inner bound relies upon. 
For any distortions their
scheme implicitly allows the receiver to perfectly
decode whatever is transmitted from transmitter-$ X $
to transmitter-$ Y $. By contrast, we do not impose
this constraint in the achievability scheme that yields
Theorem~\ref{achiev}. More generally, by relaxing this
constraint we obtain an achievable rate region that
contains, and in certain cases strictly, the rate region given
by  \cite[Theorem 5.1]{KaspBerg82}. This is given by Theorem~\ref{achiev-rd1} hereafter. For the
specific
full cooperation case, Theorem~\ref{achiev-rd1} reduces to
\cite[Theorems 5.4]{KaspBerg82}. As a result,
Theorem~\ref{achiev-rd1} always includes the convex
hull of the two regions \cite[Theorems 5.1 and
5.4]{KaspBerg82} generalized to arbitrary functions. Moreover, this inclusion is strict in
certain cases as shown through Examples~\ref{ex:KaspiBerger1} and \ref{ex:KaspiBerger2}.

\begin{theorem}[Inner Bound -- Rate Distortion] \label{achiev-rd1} $(R_0,R_X,R_Y) $ is achievable  with distortions $D_1$ and $D_2$ whenever
\begin{align} R_0 &> I(X;U|Y) \nonumber \\ 
R_X &> I(V;X|T,W) \nonumber \\  
R_Y &> I(U,Y;W|V,T) \nonumber \\  
R_X+R_Y &> I(X,Y;V,T,W)+I(U;W|V,X,T,Y)\nonumber 
\end{align} for some $ T $, $ U $, $V$, and $W$ that satisfy the Markov conditions
$$T-U-X-Y$$ $$V - (X,T) - (U,Y)-W,$$ 
and for which there exists two functions $ g_1(V,T,W) $ and $
g_2(V,T,W) $ such that
$$ \mathbb{E}d_i(f_i(X,Y),g_i(V,T,W)) \leq D_i, i\in
\{1,2\}. $$
Moreover, the following cardinality bounds hold
\begin{align*}
|\mathcal{T}| & \leq |\mathcal{X}|+4 \\
|\mathcal{V}| & \leq (|\mathcal{X}|+4)\cdot |\mathcal{X}|+1 \\
|\mathcal{W}| & \leq |\mathcal{U}| \cdot |\mathcal{Y}|+1.
\end{align*}
on the alphabets $\mathcal{T}$, $\mathcal{U}$, $\mathcal{V}$, $\mathcal{W}$, of  $ T $, $ U $, $V$, $W$, respectively.

\end{theorem}

To obtain the general inner bound
\cite[Theorem 5.1]{KaspBerg82}  it suffices to let $ T=U $
in Theorem~\ref{achiev-rd1}. To obtain the specific full
cooperation inner bound \cite[Theorem
5.4]{KaspBerg82}, it suffices
to let $ U=X $ and let $ V $  be a constant in
Theorem~\ref{achiev-rd1}. Hence, Theorem~\ref{achiev-rd1} always includes the
 convex hull of the two schemes
 \cite[Theorems 5.1 and 5.4]{KaspBerg82}. The following two examples show that 
 this inclusion can be strict. 
 
 In the first example one of the distortion functions is defined on both sources $X$ and $Y$, while in the second example the distortion functions are defined on each sources separately, as considered by Kaspi and Berger (see \cite[Section II]{KaspBerg82}).

\begin{example} \label{ex:KaspiBerger1}

Let $(X=(X_1,X_2),Y)$ where $ X_1 $
and $ Y $ are uniformly distributed over $ \{1,2,3\} $ and $ X_2$ is
a Bernoulli($p$) random variable with $p \leq
\frac{1}{2} $. Random variables $X_1$, $X_2$, and $Y$ are supposed to be jointly independent. Define the binary function $ f(X_1,Y)$ to
be equal to $ 1 $ whenever $ X_1=Y $ and equal to $ 0 $ otherwise.
The goal is to reconstruct  $f(X_1,Y)$ with average
Hamming distortion equal to zero ({\it{i.e.}}, $D_1=0$) and $X_2$ with
average Hamming distortion  $D_2\leq p$. 

For any value of $ R_0 $, the achievable scheme \cite[Theorem 5.1]{KaspBerg82}
gives
\begin{align}
R_X+R_Y > H(X_1)+ H_b(p)-H_b(d). \label{eq:sum5.1}
\end{align}
To see this note that the achievable scheme that
yields  \cite[Theorem 5.1]{KaspBerg82} is so that
whatever transmitter-$X$ sends to transmitter-$Y$ will be retransmitted to the receiver.
Therefore,  the sum rate is at least as large
as the point-to-point rate distortion problem where the
transmitter has access to $ X $ and the receiver, who has access to $ Y $,
wants to recover $ f(X_1,Y) $ and $ X_2 $ with
distortions $ 0 $ and $ D_2 $, respectively. For
the point-to-point case, due to the independence of
$(X_1,Y)$ and $X_2 $, the infimum of sum rate is at least
$$R_0(f(X_1,Y))+R_2.$$ 
Here $ R_0(f(X_1,Y)) $ is the infimum of number of bits for
recovering $ f(X_1,Y) $ with zero distortion, which is
equal to  $H(X_1)$ due to \cite[Theorem
2]{OrliRoc01}, and $R_2$ is the infimum of number of
bits for recovering $ X_2 $ with distortion $ D_2 \leq p $
and is equal to $H_b(p)-H_b(d)$ by \cite[Theorem 10.3.1]{CoverTho06}. 
Inequality \eqref{eq:sum5.1} then follows.

Now, for the scheme  \cite[Theorem 5.4]{KaspBerg82} the infimum of sum rate for $ R_0>H(X|Y) $ is
\begin{align}
R_X+R_Y = H(f(X_1,Y))+ H_b(p)-H_b(d). \label{eq:sum5.4}
\end{align}

Therefore, from \eqref{eq:sum5.1} and \eqref{eq:sum5.4}
the time sharing of \cite[Theorems
5.1]{KaspBerg82} and \cite[Theorems 5.4]{KaspBerg82} gives
\begin{align}
R_X+R_Y > q \cdot H(f(X_1,Y))&+(1-q)\cdot H(X_1) + H_b(p)-H_b(d) \label{eq:sum5}
\end{align}
for $q\in [0,1]$.
To have an  average cooperation at most equal to $ R_0
$, the time-sharing constant $q$ should be less than $\frac{R_0}{H(X|Y)} $. This is because the scheme
\cite[Theorem 5.4]{KaspBerg82} needs, on average, more than $
H(X|Y) $ cooperation bits. Now, since $ H(f(X_1,Y))<H(X_1)$, the larger $q$ is the smaller the 
right-hand side of \eqref{eq:sum5}, and therefore
\begin{align}\label{eq:sum18}
R_X+R_Y \geq \frac{R_0}{H(X|Y)}\cdot H(f(X_1,Y))+\left(1-\frac{R_0}{H(X|Y)}\right) \cdot H(X_1) + H_b(p)-H_b(d)\,.
\end{align}

We now turn to  Theorem~\ref{achiev-rd1}. By letting $
U=X_1 $, $ T $ be a constant, $ W=f(X_1,Y) $, and $ V=\text{Bernoulli}(\frac{p-d}{1-2  d}) $ be an input to the binary symmetric channel with parameter $d$ that yields $X_2$ such that\footnote{We use $\oplus$ to denote the sum modulo 2.} 
$$p_{V|X,Y}(v|x,y)=p_{V|X_2}(v|x_2)=\frac{p_Z(x_2 \oplus v)\cdot p_V(v)}{p_{X_2}(x_2)},$$
 Theorem~\ref{achiev-rd1} gives
  for $ R_0>H(X_1|Y) $ the sum rate 
\begin{align}
R_X+R_Y = H(f(X_1,Y))+ H_b(p)-H_b(d), \label{eq:sumTh}
\end{align}
which can be checked to be strictly below the right-hand
side of \eqref{eq:sum18} for $ H(X_1|Y) < R_0 < H(X|Y)$.
\end{example}

\begin{example} \label{ex:KaspiBerger2}
Let $X$ and $Y$ be random variables taking values in $\{-1,0,+1\}$ with probabilities
\begin{gather}
\raisetag{-10pt}
p(x,y) = \begin{cases}
0 \mbox{  if } (x,y)=(-1,+1) \mbox{ or } (x,y)=(+1,-1),\\
\frac{1}{7}  \mbox{ otherwise.} 
\end{cases} \nonumber
\end{gather}
Define the distortion function

\begin{gather}
\raisetag{-10pt}
d_1(x,\hat{x}) = \begin{cases}
1 \mbox{  if } x \cdot \text{sign}(\hat{x})=-1,\\
0  \mbox{ otherwise.} 
\end{cases} \nonumber
\end{gather}
where 
\begin{gather}
\raisetag{-10pt}
\text{sign}(\hat{x}) = \begin{cases}
+1 \mbox{  if } \hat{x}>0,\\
-1 \mbox{  if } \hat{x}<0,\\
0  \mbox{  if } \hat{x}=0,
\end{cases} \nonumber
\end{gather}
Let $d_2(\cdot,\cdot)=d_1(\cdot,\cdot)$. We consider the rate region for the distortion pair $(D_1,D_2)=(0,0)$. 

We claim that 
\begin{itemize}
\item[1. ] for any value of $ R_0 $ in \cite[Theorem 5.1]{KaspBerg82}
\begin{align}
R_X+R_Y > 1.03, \label{eq:ex2sum5.1}
\end{align}
\item[2. ] the infimum of the sum rate in \cite[Theorem 5.4]{KaspBerg82} under full cooperation $ R_0>H(X|Y)=1.25 $ is
\begin{align}
R_X+R_Y = 0.85, \label{eq:ex2sum5.4}
\end{align}
\item[3. ] from Theorem~\ref{achiev-rd1} it is possible to achieve the sum rate 
\begin{align}
R_X+R_Y = 0.85 \label{eq:ex2sum5.4}
\end{align}
for any $R_0>0.38$.
\end{itemize}
From 1. and 2. it can be concluded that any time sharing of the schemes \cite[Theorem 5.1]{KaspBerg82} and \cite[Theorem 5.4]{KaspBerg82} that achieves $R_0=0.39$, yields a sum rate larger than $0.89$, which is larger than the sum rate given by Theorem~\ref{achiev-rd1}.

The proofs of Claims 1.-3. are deferred to the Appendix.
\end{example}

\section{Analysis} \label{sec:analysis}
\begin{IEEEproof}[Proof of Theorem~\ref{achiev}]
Pick  $ T $, $ U $, $V$, and $W$ as in the theorem. 
These random variables together with $(X,Y)$ are
distributed according to some joint probability $p(v,x,t,u,y,w)$. 

The coding procedure consists of two phases. In the first phase transmitter-$X$ sends $ (\mathbf{T}(\mathbf{X}),\mathbf{U}(\mathbf{T}(\mathbf{X})))$ to transmitter-$Y$. In the 
second phase, both transmitters send $\mathbf{T}(\mathbf{X})$ to the receiver. In addition to this message, transmitter-$ X $ and transmitter-$ Y $ send 
$\mathbf{V}(\mathbf{X},\mathbf{T}(\mathbf{X}))) $ and $ \mathbf{W}(\mathbf{Y},\mathbf{T}(\mathbf{X}),\mathbf{U}(\mathbf{T}(\mathbf{X}))) $,
respectively, to the receiver. As can be seen, only part of the message sent from transmitter-$ X $ to transmitter-$ Y $, $\mathbf{T}(\mathbf{X})$, is retransmitted from both transmitters to the receiver while for the other part, $\mathbf{U}(\mathbf{T}(\mathbf{X}))$, a function of it $ \mathbf{W}(\mathbf{Y},\mathbf{T}(\mathbf{X}),\mathbf{U}(\mathbf{T}(\mathbf{X}))) $ is sent by transmitter-$Y$ to the receiver. Details follow.

For $ t \in \mathcal{T} $, ${v} \in \Gamma(G_{T,X|T,U,Y})$, and ${w} \in
\Gamma(G_{T,U,Y|T,V})$, define 
$\tilde{f}(v,t,w)$ to be equal to
$f(x,y)$ for all $(t,x) \in {v}$ and  $ (t,u,y) \in
{w}$ such that $p(x,t,u,y)>0$. Further, for
$\mathbf{t}=(t_{1},\ldots,t_{n})$,
$\mathbf{v}=(v_{1},\ldots,v_{n})$, and
$\mathbf{w}=(w_{1},\ldots,w_{n})$ let 
$$\tilde{f}(\mathbf{v},\mathbf{t},\mathbf{w})\defeq\tilde{f}(v_{1},t_{1},w_{1}),\ldots,\tilde{f}(v_{n},t_{n},w_{n}) \,.$$

Generate $2^{nR_{T}}$ sequences
$$\mathbf{t}^{(i)}=(t^{(i)}_{1},t^{(i)}_{2},\ldots,t^{(i)}_{n})\,,$$
$i\in \{1,2,\ldots,2^{nR_{T}}\}$, i.i.d.
according to the marginal distribution $p(t)$.

For each codeword $\mathbf{t}^{(i)}$, generate $2^{nR_{U}}$ sequences
$$\mathbf{u}^{(j)}(\mathbf{t}^{(i)})=(u^{(j)}_{1}(t_1^{(i)}),u^{(j)}_{2}(t_2^{(i)}),\ldots,u^{(j)}_{n}(t_n^{(i)}))\,,$$
$j\in \{1,2,\ldots,2^{nR_{U}}\}$, i.i.d.
according to the marginal distribution $p(u|t)$, and randomly bin each 
sequence $ (\mathbf{t}^{(i)},\mathbf{u}^{(j)}(\mathbf{t}^{(i)})) $ uniformly into $2^{nR_0}$ bins. Similarly, generate
$2^{nR_{V}}$ and $2^{nR_{W}}$ sequences
$$\mathbf{v}^{(k)}(\mathbf{t}^{(i)})=(v^{(k)}_1(t_1^{(i)}),v^{(k)}_2(t_2^{(i)}),\ldots,v_n^{(k)}(t_n^{(i)})),$$ 
and
$$\mathbf{w}^{(l)}(\mathbf{t}^{(i)})=(w^{(l)}_{1}(t_1^{(i)}),w^{(l)}_{2}(t_2^{(i)}),\ldots,w^{(l)}_{n}(t_n^{(i)})),$$ 
respectively, i.i.d. according
to $p(v|t)$ and $ p(w|t) $, respectively, and randomly and uniformly bin each 
sequence $ (\mathbf{t}^{(i)},\mathbf{v}^{(k)}(\mathbf{t}^{(i)})) $ and 
$ (\mathbf{t}^{(i)},\mathbf{w}^{(l)}(\mathbf{t}^{(i)})) $ into
$2^{nR_X}$ and $2^{nR_Y}$ bins, respectively. Reveal the bin 
assignment $\phi_0$  to the both encoders and the bin assignments
$\phi_X$ and $\phi_Y$ to the encoders and the decoder.

Let $\varepsilon'$ and $\varepsilon''$ be two constants such that $\varepsilon'>\varepsilon''>0$.

\noindent\textit{Encoding} 

First phase: Transmitter-$ X $
tries to find a sequence
$(\mathbf{t},\mathbf{u}(\mathbf{t}))$ that is 
 jointly typical with
$\mathbf{x}$, \textit{i.e.},\footnote{$\mathcal{T}_{\varepsilon}^{(n)}(X,Y)$ is the set of jointly $\varepsilon$-typical $n$-sequences.}  $(\mathbf{t},\mathbf{u}(\mathbf{t}),\mathbf{x}) \in \mathcal{T}_{\varepsilon''}^{(n)}(T,U,X)$ and sends the index of the bin that contains 
this sequence, {\it{i.e.}},
$\phi_0(\mathbf{t},\mathbf{u}(\mathbf{t}))\defeq q_0$, to transmitter-$ Y $.

Second phase: 
Transmitter-$ X $ tries to find a unique $\mathbf {v}(\mathbf{t})$
that is  jointly typical with
$(\mathbf{x},\mathbf{t})$, \textit{i.e.}, $(\mathbf {v}(\mathbf{t}),\mathbf{x},\mathbf{t}) \in \mathcal{T}_{\varepsilon'}^{(n)}(V,X,T)$ and sends the index of the bin that contains $
(\mathbf{t},\mathbf{v}(\mathbf{t}))$, {\it{i.e.}},
$\phi_X(\mathbf{t},\mathbf{v}(\mathbf{t}))\defeq q_X$, to the receiver.

Transmitter-$Y$ upon receiving the index $ q_0 $, 
first tries to find a unique $(\check{\mathbf {t}},\check{\mathbf {u}}(\check{\mathbf {t}}))$ 
such that $ (\check{\mathbf {t}},\check{\mathbf {u}}(\check{\mathbf
{t}}),\mathbf{y})\in \mathcal{T}_{\varepsilon'}^{(n)}(T,U,Y)$ and such that $ \phi_0(\check{\mathbf
{t}},\check{\mathbf {u}}(\check{\mathbf {t}}))=q_0 $. Then, it
tries to find a unique $\mathbf{w}(\check{\mathbf {t}})$ that is 
 jointly typical with
$ (\check{\mathbf {u}}(\check{\mathbf {t}}),\mathbf{y}) $, \textit{i.e.}, $ (\mathbf{w}(\check{\mathbf {t}}),\check{\mathbf {u}}(\check{\mathbf {t}}),\mathbf{y}) \in \mathcal{T}_{\varepsilon'}^{(n)}(W,U,Y)$ and sends the index of the bin that contains 
$(\check{\mathbf {t}},\mathbf{w}(\check{\mathbf {t}}))$, {\it{i.e.}},
$q_Y=\phi_Y(\check{\mathbf {t}},\mathbf{w}(\check{\mathbf {t}}))$, to the receiver.

If a transmitter cannot find an index as above, it declares an error, and if there is more than one index, the transmitter selects one of them randomly and
uniformly.

\noindent\textit{Decoding:} Given  the index pair
$(q_X,q_Y)$, declare
$\tilde{f}(\hat{\mathbf{v}}(\hat{\mathbf{t}}),{\hat{\mathbf{t}},\hat{\mathbf{w}}}(\hat{\mathbf{t}}))$ if there
exists a unique jointly typical
$(\hat{\mathbf{v}},{\hat{\mathbf{t}},\hat{\mathbf{w}}})\in \mathcal{T}_{\varepsilon}^{(n)}(V,T,W)$ such that
$\phi_X(\hat{\mathbf{t}},\hat{\mathbf{v}}(\hat{\mathbf{t}}))=q_X$ and
$\phi_Y(\hat{\mathbf{t}},\hat{\mathbf{w}}(\hat{\mathbf{t}}))=q_Y$, and such that
$\tilde{f}(\hat{\mathbf{v}}(\hat{\mathbf{t}}),{\hat{\mathbf{t}},\hat{\mathbf{w}}}(\hat{\mathbf{t}}))$
is defined. Otherwise declare an
error.

\noindent \textit{Probability of Error:} In each of the two phases there are two types of error.

{\textit{First phase:}} in the first phase,  the first type of error occurs
when  no $(\mathbf{t},\mathbf{u}(\mathbf{t}))$ is jointly typical with 
$\mathbf{x}$. The
probability of this error is negligible for $n$ large enough, due to the covering lemma \cite[Lemma~3.3]{ElgaKim12} whenever
\begin{align}
R_T &\geq I(X;T)+\delta_1(\varepsilon''), \nonumber \\
R_U &\geq I(X;U|T)+\delta_2(\varepsilon''),  \label{eq:sizeOfTU}
\end{align}
where $\delta_1(\varepsilon'')$ and $\delta_2(\varepsilon'')$ tend to zero as $\varepsilon''$ tends to zero.

The second type of error occurs if $ (\check{\mathbf{t}},\check{\mathbf{u}}(\check{\mathbf{t})}) \neq (\mathbf{t},\mathbf{u}(\mathbf{t}))$. By symmetry of the scheme, this error probability, is the same as the average
error probability conditioned on the transmitter-$X$ selecting
$ \mathbf{T}^{(1)} $ and $\mathbf{U}^{(1)}(\mathbf{T}^{(1)})$. So, we consider the error event
\begin{align}
\mathcal{E}'\defeq \{(\hat{\mathbf{T}},\hat{\mathbf{U}}(\hat{\mathbf{T}}))&\ne(\mathbf{T}^{(1)},\mathbf{U}^{(1)}(\mathbf{T}^{(1)}))\}.  \label{eq:achNeqTU}
\end{align}
Define the following events
\begin{align}
\mathcal{E}'_{i,j}\defeq \{(\mathbf{T}^{(i)},\mathbf{U}^{(j)}(\mathbf{T}^{(i)})) \in \mathcal{T}_{\varepsilon'}^{(n)}(T,U,Y), \phi_0 (\mathbf{T}^{(i)},\mathbf{U}^{(j)}(\mathbf{T}^{(i)}))= q_0
\}.  \nonumber
\end{align}
Hence we have
\begin{align}
P(\mathcal{E}')=& P\big(\mathcal{E}_{1,1}^{'c} \cup (\bigcup_{j \neq 1} \mathcal{E}'_{1,j} )\cup  (\bigcup_{i \neq 1} \mathcal{E}'_{i,1}) \cup (\bigcup_{i \neq 1,j \neq 1} \mathcal{E}'_{i,j})\big) \nonumber \\
\leq &P(\mathcal{E}_{1,1}^{'c}) + \sum \limits_{j \neq 1} P(\mathcal{E}'_{1,j}) + \sum \limits_{i \neq 1} P(\mathcal{E}'_{i,1})+ \sum \limits_{i \neq 1,j \neq 1} P(\mathcal{E}'_{i,j}). \label{eq:proofAchievTU1}
\end{align}

According to the properties of jointly 
 typical sequences (See \cite[Section~2.5]{ElgaKim12}), for any $\varepsilon'>\varepsilon''>0$ we have
\begin{itemize}
\item $P(\mathcal{E}_{1,1}^{'c}) \leq \delta'(\varepsilon',\varepsilon'')$ due to the encoding process and the Markov chain $T-U-X-Y$;
\item for $j \neq 1$, $$P(\mathcal{E}'_{1,j}) \leq 2^{-n(I(U;Y|T)-\delta'_1(\varepsilon'))}\cdot 2^{-nR_0};$$
\item for $i \neq 1$, $$P(\mathcal{E}'_{i,1}) \leq 2^{-n(I(T,U;Y)-\delta'_2(\varepsilon'))}\cdot 2^{-nR_0};$$
\item for $i \neq 1$ and $j \neq 1$, $$P(\mathcal{E}'_{i,j}) \leq 2^{-n(I(T,U;Y)-\delta'_3(\varepsilon')))}\cdot 2^{-nR_0};$$
\end{itemize}
where $\delta'(\varepsilon',\varepsilon'')$, $\delta'_1(\varepsilon')$, $\delta'_2(\varepsilon')$, and  $\delta'_3(\varepsilon')$ tend to zero as $\varepsilon'$ tends to zero.

Using the above bounds, the probability of error in \eqref{eq:proofAchievTU1} can be bounded as
\begin{align}
P(\mathcal{E}')\leq &  \delta'(\varepsilon',\varepsilon'') +2^{nR_T} \times 2^{-n(I(U;Y|T)-\delta'_1(\varepsilon'))}\cdot 2^{-nR_0}\nonumber \\
&+2^{nR_T} \cdot 2^{-n(I(T,U;Y)-\delta'_2(\varepsilon'))}\cdot 2^{-nR_0}\nonumber \\
&+2^{n(R_T+R_U)} \cdot 2^{-n(I(T,U;Y)-\delta'_3(\varepsilon')))}\cdot 2^{-nR_0}\nonumber
\end{align}
Hence, using \eqref{eq:sizeOfTU}, the error probability can be made to vanish whenever $n$ tends to infinity as long as  
\begin{align}
R_0 > I(X;U,T)-I(U,T;Y)=I(X;U,T|Y)=I(X;U|Y), \label{eq:achR0}
\end{align}
where the equalities are due to the Markov chain $T-U-X-Y$.

{\textit{Second phase:}} in the second phase, the first type of error occurs
when  no $\mathbf{v}(\mathbf{t})$, respectively no $\mathbf{w}(\check{\mathbf{t}})$,  is 
jointly typical with $(\mathbf{x},\mathbf{t})$,  respectively with $(\check{\mathbf{u}}(\check{\mathbf{t}}),\mathbf{y})$. Due to the covering lemma \cite[Lemma~3.3]{ElgaKim12} the
probability of each of these two errors is negligible for $n$ large enough whenever
\begin{align}
R_V &\geq I(V;X|T)+\delta'_4(\varepsilon'), \nonumber \\
R_U &\geq I(U,Y;W|T)+\delta'_5(\varepsilon'),  \label{eq:sizeOfVW}
\end{align}
where $\delta'_4(\varepsilon')$ and $\delta'_5(\varepsilon')$ tend to zero as $\varepsilon'$ tends to zero.

The second type of error refers to the
Slepian-Wolf coding procedure. By symmetry of the scheme, the average error probability of 
the Slepian-Wolf coding procedure, is the same as the average
error probability conditioned on the transmitters selecting
$ \mathbf{T}^{(1)} $, $\mathbf{U}^{(1)}(\mathbf{T}^{(1)})$
,$\mathbf{V}^{(1)}(\mathbf{T}^{(1)})$ and
$\mathbf{W}^{(1)}(\mathbf{T}^{(1)})$. Note that since $\mathbf{t}$, $\mathbf{u}(\mathbf{t})$, $\mathbf{v}(\mathbf{t})$, and $\mathbf{w}(\mathbf{t})$ are chosen such that  
$$(\mathbf{t},\mathbf{u}(\mathbf{t}),\mathbf{x}) \in \mathcal{T}_{\varepsilon''}^{(n)}(T,U,X)$$
$$(\mathbf {v}(\mathbf{t}),\mathbf{x},\mathbf{t}) \in \mathcal{T}_{\varepsilon'}^{(n)}(V,X,T)$$ $$ (\mathbf{w}(\check{\mathbf {t}}),\check{\mathbf {u}}(\check{\mathbf {t}}),\mathbf{y}) \in \mathcal{T}_{\varepsilon'}^{(n)}(W,U,Y),$$ by the definition of jointly typical sequences \cite[Section~2.5]{ElgaKim12} and the Markov chains~\eqref{eq:th1Markovs} we have $p(t_i,x_i,v_i)>0$ and $p(t_i,u_i,y_i,w_i)>0$ for any $1 \leq i \leq n$. Hence we have $(t_i,x_i) \in v_i$ and $(t_i,u_i,y_i) \in w_i$. Combining this with the definitions of 
 $T$, $V$, $W$ and $ \tilde{f}(V,T,W) $ implies that if the transmitted messages are decoded
 correctly, then the function is computed without error.
 
 We now consider the error event
\begin{align}
\mathcal{E}\defeq \{(\hat{\mathbf{T}},\hat{\mathbf{V}}(\hat{\mathbf{T}}),\hat{\mathbf{W}}(\hat{\mathbf{T}}))&\ne(\mathbf{T}^{(1)},\mathbf{V}^{(1)}(\mathbf{T}^{(1)})
,\mathbf{W}^{(1)}(\mathbf{T}^{(1)}))\}.  \label{eq:achNeq}
\end{align}
and assume that the transmitters selected
$ \mathbf{T}^{(1)} $, $\mathbf{U}^{(1)}(\mathbf{T}^{(1)})$
,$\mathbf{V}^{(1)}(\mathbf{T}^{(1)})$ and
$\mathbf{W}^{(1)}(\mathbf{T}^{(1)})$.

Define the following events,
\begin{align}
\mathcal{E}_{i,k,l}\defeq \{&(\mathbf{T}^{(i)},\mathbf{V}^{(k)}(\mathbf{T}^{(i)})
,\mathbf{W}^{(l)}(\mathbf{T}^{(i)})) \in \mathcal{T}_{\varepsilon}^{(n)}(T,V,W), \nonumber \\
&(\phi_X (\mathbf{T}^{(i)},\mathbf{V}^{(k)}(\mathbf{T}^{(i)})),\phi_Y (\mathbf{T}^{(i)},\mathbf{W}^{(l)}(\mathbf{T}^{(i)}))= (q_X,q_Y)
\}.  \nonumber
\end{align}
We have
\begin{align}
P(\mathcal{E})=& P(\mathcal{E}_{1,1,1}^c \cup (\bigcup_{k \neq 1} \mathcal{E}_{1,k,1} )\cup  (\bigcup_{l \neq 1} \mathcal{E}_{1,1,l}) \cup (\bigcup_{k \neq 1,l \neq 1} \mathcal{E}_{1,k,l}) \cup (\bigcup_{i \neq 1,k,l} \mathcal{E}_{i,k,l})) \nonumber \\
\leq &P(\mathcal{E}_{1,1,1}^c) + \sum \limits_{k \neq 1} P(\mathcal{E}_{1,k,1}) + \sum \limits_{l \neq 1} P(\mathcal{E}_{1,1,l})+ \sum \limits_{k \neq 1,l \neq 1} P(\mathcal{E}_{1,k,l}) + \sum \limits_{i \neq 1,k,l} P(\mathcal{E}_{i,k,l}). \label{eq:proofAchiev1}
\end{align}

According to the properties of jointly 
 typical sequences (See \cite[Section~2.5]{ElgaKim12}), for any $\varepsilon>\varepsilon'>0$ we have
\begin{itemize}
\item $P(\mathcal{E}_{1,1,1}^c) \leq \delta(\varepsilon,\varepsilon')$ due to the encoding process and the Markov chain $V-(T,X)-(U,Y)-W$;
\item for $k \neq 1$, $$P(\mathcal{E}_{1,k,1}) \leq 2^{-n(I(V;W|T)-\delta_1(\varepsilon))}\cdot 2^{-nR_X};$$
\item for $l \neq 1$, $$P(\mathcal{E}_{1,1,l}) \leq 2^{-n(I(V;W|T)-\delta_2(\varepsilon))}\cdot 2^{-nR_Y};$$
\item for $k \neq 1$, $l \neq 1$, $$P(\mathcal{E}_{1,k,l}) \leq 2^{-n(I(V;W|T)-\delta_3(\varepsilon)))}\cdot 2^{-nR_X}\cdot 2^{-nR_Y};$$
\item for $i \neq 1$, $$P(\mathcal{E}_{i,k,l}) \leq 2^{-n(I(V;W|T)-\delta_4(\varepsilon))}\cdot 2^{-nR_X}\cdot 2^{-nR_Y};$$
\end{itemize}
where $\delta(\varepsilon,\varepsilon')$, $\delta_1(\varepsilon)$, $\delta_2(\varepsilon)$, $\delta_3(\varepsilon)$, and $\delta_4(\varepsilon)$ tend to zero as $\varepsilon$ tends to zero.

Using the above bounds, the probability of error in \eqref{eq:proofAchiev1} can be bounded as
\begin{align}
P(\mathcal{E})\leq &  \delta(\varepsilon,\varepsilon') +2^{nR_V} \cdot 2^{-n(I(V;W|T)-\delta_1(\varepsilon))}\cdot 2^{-nR_X}\nonumber \\
&+2^{nR_W} \cdot 2^{-n(I(V;W|T)-\delta_2(\varepsilon))}\cdot 2^{-nR_Y}\nonumber \\
&+2^{nR_V} \cdot 2^{nR_W} \cdot 2^{-n(I(V;W|T)-\delta_3(\varepsilon))}\cdot 2^{-nR_X}\cdot 2^{-nR_Y}\nonumber \\
&+2^{nR_T} \cdot 2^{nR_V} \cdot 2^{nR_W} \cdot 2^{-n(I(V;W|T)-\delta_4(\varepsilon))}\cdot 2^{-nR_X}\cdot 2^{-nR_Y}.\nonumber
\end{align}
Combining this inequality with \eqref{eq:sizeOfTU}, \eqref{eq:achR0}, and \eqref{eq:sizeOfVW} implies that the error probability goes to zero whenever $n$ goes to infinity and inequalities \eqref{eq:th1Rates} are satisfied.

We now show that the rate region is maximal under under the cardinality bounds \eqref{eq:th1Cardinalities}. Suppose $(V,X,T,U,Y,W)\sim p(v,x,t,u,y,w)$ satisfy \eqref{eq:th1Markovs} and \eqref{eq:th1Sets}. 

\noindent \textbf{Cardinality of $\mathcal{T}$}:  Suppose that $|\mathcal{T}| \geq |\mathcal{X}|+5$. We replace random variable $T$ by a new random variable $T'\in {\mathcal{T'}} $ such that ${\mathcal{T'}}\subseteq {\mathcal{T}}$ and $|{\mathcal{T'}}|\leq |\mathcal{X}|+4$. Random variables $(V,X,T',U,Y,W)$ are thus distributed according to some joint probability $p'(v,x,t,u,y,w)$,  defined as
\begin{align}\label{ppprime}
p'(v,x,t,u,y,w)=p(v,x,u,y,w|t)p'(t).
\end{align}
Distribution $p'(v,x,t,u,y,w)$ 
\begin{itemize}
\item  is admissible in the sense that it satisfies \eqref{eq:th1Markovs} and \eqref{eq:th1Sets},
\item  and achieves the same right-hand side terms in inequalities \eqref{eq:th1Rates} as $p(v,x,t,u,y,w)$.
\end{itemize}
Condition \eqref{eq:th1Sets} holds since ${\mathcal{T}}\subseteq {\mathcal{T}'}$. 

To show that \eqref{eq:th1Markovs} are satisfied by $p'(v,x,t,u,y,w)$ regardless of the choice of $p'(t)$, we show that $U-X-Y$ and $X-(U,Y)-W$ also hold under $p'(v,x,t,u,y,w)$. We have
\begin{align*}
p'(x,y|u)&=\sum \limits_t p'(x,y|u,t) \cdot p'(t|u) \\
&\stackrel{(a)}{=}  \sum \limits_t p(x,y|u,t) \cdot p'(t|u)\\
&\stackrel{(b)}{=} \sum \limits_t p(x,y|u) \cdot p'(t|u)= p(x,y|u) \cdot \sum \limits_t p'(t|u)\\
&=p(x,y|u)
\end{align*}
where $(a)$ follows from the fact that $p'(u,x,y|t)=p(u,x,y|t)$ and where $(b)$ follows from the Markov chain $T-U-X-Y$. By a similar computation, one shows that concludes that $p'(w|y,u,x)=p'(w|y,u,x)$.

The fact that $p'(x,y,u|t)=p(x,y,u|t)$ and $p'(x,y|u)=p(x,y|u)$ imply that the Markov chain $T'-U-X-Y$ hold for any distribution $p'(t), t \in \mathcal{T}$. Similarly, since $p'(v,x,u,y,w|t)=p(v,x,u,y,w|t)$ and $p'(w|y,u,x)=p(w|y,u,x)$, the Markov chain $V-(X,T')-(U,Y)-W$ also holds for any $p'(t)$. We thus established that $p'(v,x,t,u,y,w)$ satisfies the Markov chains \eqref{eq:th1Markovs} for any marginal $p'(t), t \in \mathcal{T}$. 

Now we construct a specific distribution $p'(t), t \in \mathcal{T}'$, such that $ |\mathcal{T}'|\leq |\mathcal{X}|+4$, and such that the right-hand sides of inequalities \eqref{eq:th1Rates} remain unchanged when $p(v,x,t,u,y,w)$ is replaced by $p'(v,x,t,u,y,w)$. 
As a distribution, $p'(t)$ should satisfy 
\begin{align}
\sum \limits_{t \in \mathcal{T}} p'(t)=1  \label{eq:cardinalityT1}.
\end{align}
Now, consider the right-hand side of the first term in \eqref{eq:th1Rates}:
\begin{align*}
I(X;U|Y)|_{p(t)}=H(X|Y)|_{p(t)}-H(X|U,Y)|_{p(t)}
\end{align*}
where we emphasized the dependency on $p(t)$ of the joint distribution $p(v,x,t,u,y,w)$.
To have $$H(X|Y)|_{p'(t)}=H(X|Y)|_{p(t)}$$ it suffices to have $p'(x)=p(x)$ (since $p'(y|x)=p(y|x)$ from \eqref{ppprime} and the first Markov condition in \eqref{eq:th1Markovs}). This condition is satisfied if 
\begin{align}
\sum \limits_{t \in \mathcal{T}'}p(x=i|t)\cdot p'(t)=p(x=i)|_{p(t)} \quad 1 \leq i \leq |\mathcal{X}|-1  \label{eq:cardinalityT2},
\end{align}
(as above $p(x=i)|_{p(t)}$ denotes the $X$-marginal under the original distribution $p(v,x,t,u,y,w)$).

To have $$H(X|U,Y)|_{p'(t)}=H(X|U,Y)|_{p(t)}$$
we impose
\begin{align}
\sum \limits_{t \in \mathcal{T}}a_t \cdot p'(t)=b, \label{eq:cardinalityT3}
\end{align}
where $$a_t\defeq \sum \limits_{u,y} H(X|U=u,Y=y)p(y|u)p(u|t)$$ and $$b\defeq H(X|U,Y)|_{p(t)}.$$ Note that the $a_t$'s and $b_t$'s do not depend on $p'(t)$.
Therefore, under \eqref{eq:cardinalityT2} and \eqref{eq:cardinalityT3} we get
$$I(X;U|Y)|_{p'(t)}=I(X;U|Y)|_{p(t)},$$ implying that the first condition in \eqref{eq:th1Rates} remains unchanged. 

Similarly, for keeping the right-hand sides of the three other inequalities in \eqref{eq:th1Rates},  $p'(t)$ should satisfy the set of linear equations
\begin{align}
\sum \limits_{t \in \mathcal{T}} I(V;X|W,T=t)\cdot p'(t)&=I(V;X|T,W)|_{p(t)}, \label{eq:cardinalityT4} \\
\sum \limits_{t \in \mathcal{T}} I(U,Y;W|V,T=t)\cdot  p'(t)&=I(U,Y;W|V,T)|_{p(t)}, \label{eq:cardinalityT5} \\
\sum \limits_{t \in \mathcal{T}} (H(X,Y|V,T=t,W)&+I(U;W|V,X,T=t,W))\cdot p'(t)\nonumber \\
&=(H(X,Y|V,T,W)+I(U;W|V,X,T,W))|_{p(t)}. \label{eq:cardinalityT6} 
\end{align}
Combining, we deduce that the distribution $p'(t)$ should satisfy the set of $$m=|\mathcal{X}|+4$$ linear equations \eqref{eq:cardinalityT1}-\eqref{eq:cardinalityT6}. We write these equations in the matrix form
 \begin{align}
 A_{n \times m} \cdot Z_{m \times 1} =B_{n \times 1}. \label{eq:cardinalityMatrix}
 \end{align}
where $n=|\mathcal{T}|(\geq |\mathcal{X}|+4)\text{ by assumption})$, where $Z$ denotes the vector of $p'(t),t \in \mathcal{T}$, where $A$ denotes the matrix of coefficients (constants on the left-hand sides of \eqref{eq:cardinalityT1}-\eqref{eq:cardinalityT6}), and where $B$ denotes the vector of constants on the right-hand side of equations  \eqref{eq:cardinalityT1}-\eqref{eq:cardinalityT6}.

 We want to find a positive solution of $Z$ in the above equation where $Z_i=0$ for at least $n-m$ indices $1 \leq i \leq n$. We find such a solution recursively by showing that if $n >m$ then there exists a solution $S$ which has at least one zero entry, say $S_i$. Then, we set $n\to n-1$, remove the corresponding column of $A$ and corresponding row of $Z$ and repeat the procedure.
 
 We know that \eqref{eq:cardinalityMatrix} has at least one non-negative solution, which is the vector $p(t)$. Therefore, if we find another solution with at least one negative entry then there also exists a solution with at least one zero entry by convexity of the space solution of \eqref{eq:cardinalityMatrix}. 

Since $n>m$, there exists a column in $A$ which is a linear combination of the other columns. Without loss of generality, suppose $A_m=\sum \limits_{i=1}^{m-1}a_i  A_i $, where $A_i$ denotes the $i$-th column of $A$. Now, if $Z=[Z_1,\cdots,Z_m]^T$ is a non-negative solution, then 
 $$Z'=[Z_1+c \cdot a_1,Z_2+c \cdot a_2,\cdots,Z_{m-1}+c \cdot a_{m-1},Z_k-c]^T,$$
 is also a solution for any value of $c$. By a suitable choice of $c$,  $Z_k-c$ can be made negative which completes the proof.

\noindent \textbf{Cardinalities of $\cal{V}$ and $\cal{W}$:} The cardinalities of $\cal{V}$ and $\cal{W}$ by $|\mathcal{T}|\times |\mathcal{X}|+1$ and $|\mathcal{U}|\times |\mathcal{Y}|+1$, respectively, follows from a standard application of Carath\'eodory's theorem.
\end{IEEEproof}

\begin{IEEEproof}[Proof of Theorem \ref{rate:partially} ]
For achievability it suffices to let $T=U$ and $V=X$ in Theorem \ref{achiev}.

Now for the converse. Let $ C_0=\varphi_0(\mathbf{X}) $ be the message received by transmitter-$Y$ and let $ C_X=\varphi_X(\mathbf{X}) $ and $ C_Y=\varphi_Y(C_0,\mathbf{Y}) $  be the received messages at the receiver from transmitter-$X$ and transmitter-$Y$ respectively. Suppose that 
 $$P(\psi(C_X,C_Y)\ne f(\mathbf{X},\mathbf{Y})) \leq \varepsilon'_n,$$
    where $\varepsilon'_n \rightarrow 0$ when $n \rightarrow \infty$. From Fano's inequality    $$H(f(\mathbf{X},\mathbf{Y})|C_X,C_Y) \leq \varepsilon_n$$
   where $\varepsilon_n \rightarrow 0$ when $n \rightarrow \infty$.

We start by  showing that the Markov chain
\begin{align}
f(X_i,Y_i)-(C_0,X_i^n,Y_1^{i-1},C_Y)-C_X \label{eq:ProofPartiallyMarkov1}
\end{align}
holds. We have
\begin{align*}
p(f(x_i,y_i)|c_0,x_i^n,y_1^{i-1},c_Y,c_X)&=\sum \limits_{x_1^{i-1}}
p(f(x_i,y_i)|c_0,x_1^n,y_1^{i-1},c_Y,c_X)\cdot p(x_1^{i-1}|c_0,x_i^n,y_1^{i-1},c_Y,c_X)\\
&\stackrel{(a)}{=}\sum \limits_{x_1^{i-1}}
p(f(x_i,y_i)|c_0,x_1^n,y_1^{i-1},c_Y)\cdot p(x_1^{i-1}|c_0,x_i^n,y_1^{i-1},c_Y,c_X) 
\end{align*}
\begin{align*}
&=\sum \limits_{x_1^{i-1}} p(x_1^{i-1}|c_0,x_i^n,y_1^{i-1},c_Y,c_X) \cdot \sum \limits_{y_i}
p(f(x_i,y_i)|x_i,y_i)\cdot p(y_i|c_0,x_1^n,y_1^{i-1},c_Y)  \\
&=\sum \limits_{x_1^{i-1}} p(x_1^{i-1}|c_0,x_i^n,y_1^{i-1},c_Y,c_X) \cdot \sum \limits_{y_i}
p(f(x_i,y_i)|x_i,y_i)\cdot p(y_i|c_0,x_i^n,y_1^{i-1},c_Y)  \\
&=\sum \limits_{x_1^{i-1}}
p(f(x_i,y_i)|c_0,x_i^n,y_1^{i-1},c_Y)\cdot p(x_1^{i-1}|c_0,x_i^n,y_1^{i-1},c_Y,c_X) \\
&=p(f(x_i,y_i)|c_0,x_i^n,y_1^{i-1},c_Y),
\end{align*}
where $(a)$ is due to the fact that $C_X$ is a function of $X_1^n$. This gives the desired Markov chain.
    
Now, by taking  $ U_i= \{C_0,X_{i+1}^n,Y_1^{i-1}\} $, $ W_i=\{C_Y,Y_1^{i-1}\} $, 
the Markov chains $ U_i-X_i-Y_i $ and $ X_i-(U_i,Y_i)-W_i $ hold and 
\begin{align*}
H(f(X_i,Y_i)|X_i,U_i,W_i)\stackrel{(a)}{=}&H(f(X_i,Y_i)|X_i,U_i,W_i,C_X) \\
\leq& H(f(X_i,Y_i)|C_Y,C_X)\\
\leq& H(f(\mathbf{X},\mathbf{Y})|C_Y,C_X) \\
\leq &\varepsilon_n,
\end{align*} 
where $(a)$ is true due to the Markov chain \eqref{eq:ProofPartiallyMarkov1}.

Then, 
\begin{align}\nonumber
nR_0 &\geq \log \limits_2|C_0| \\
&\geq H(C_0) \nonumber \\ 
& \geq I(C_0;X_1^n|Y_1^n) \nonumber \\
& \geq  \sum \limits_{i=1}^n H(X_i|Y_i)-H(X_i|Y_1^{i-1},C_0,X_{i+1}^n,Y_i)] \nonumber \\
&=\sum \limits_{i=1}^n I(X_i;U_i|Y_i) \label{eq:proofPartiallyIneq1}
\end{align}
and
\begin{align}
nR_X &\geq \log \limits_2|C_X|\nonumber  \\
&\geq H(C_X) \nonumber \\ 
& \geq I(C_X;X_1^n|C_0,C_Y) \nonumber \\
&\stackrel{(a)}{\geq}H(X_1^n|C_0,C_Y)- \varepsilon \nonumber \\
&\geq \sum \limits_{i=1}^n H(X_i|X_{i+1}^{n},C_0,C_Y,Y_{1}^{i-1})-\varepsilon \nonumber \\
&= \sum\limits_{i=1}^n  H(X_i|U_i,W_i)-\varepsilon \label{eq:proofPartiallyIneq2}
\end{align}
where $ (a) $ holds since $ \mathbf{X} $ can be 
recovered from $ (C_X,C_Y) $ with high probability since $ (C_X,C_Y) $ reveals $f(X,Y)$ with high probability and since $f(X,Y)$ is partially invertible with respect to $X$. Further,
\begin{align}
nR_Y &\geq \log \limits_2|C_Y|\nonumber \\
&\geq H(C_Y) \nonumber \\ 
& \geq I(C_Y;Y_1^n|C_0,X_1^n) \nonumber \\
&= \sum \limits_{i=1}^n [H(Y_i|X_{i+1}^n,C_0,Y_1^{i-1})-H(Y_i|X_{i+1}^n,C_0,Y_1^{i-1},C_Y,X_i)] \nonumber \\
&=\sum \limits_{i=1}^n I(Y_i;W_i|X_i,U_i)\,, \label{eq:proofPartiallyIneq3}
\end{align}
and
\begin{align}
n(R_X&+R_Y) \geq \log \limits_2 |(C_X,C_Y)| \geq H(C_X,C_Y) \nonumber \\ 
& \geq I(C_X,C_Y;X_1^n,Y_1^n) \nonumber \\
&\stackrel{(a)}{\geq} H(X_1^n)+H(Y_1^n|X_1^n)-H(Y_1^n|X_1^n,C_X,C_Y)- \varepsilon \nonumber \\
&= H(X_1^n)+H(Y_1^n|X_1^n,C_0)-H(Y_1^n|X_1^n,C_0,C_X,C_Y)- \varepsilon \nonumber \\
&\geq \sum \limits_{i=1}^n [H(X_i)+H(Y_i|X_{i+1}^n,C_0,Y_1^{i-1},X_i)-H(Y_i|Y_1^{i-1},X_{i+1}^n,C_0,C_Y,X_i)] -\varepsilon \nonumber \\
&= \sum \limits_{i=1}^n [H(X_i)+I(Y_i,W_i|X_i,U_i)] -\varepsilon \label{eq:proofPartiallyIneq4} 
\end{align}
where $ (a) $ comes from the fact that $ \mathbf{X} $ can be recovered with high probability knowing $ (C_X,C_Y) $.

Let $Q$ be a uniform random variable over $\{1,2,\cdots,n\}$. Let
\begin{align*}
X&\defeq X_Q \\
Y&\defeq Y_Q \\
U&\defeq (U_Q,Q)\\
W&\defeq (W_Q,Q).
\end{align*} 
Note that by knowing $U$ or $W$, one knows $Q$. 

In the remaining part of the proof, we first show that $U$ and $W$ satisfy the inequalities of the theorem, the Markov chains \eqref{eq:rateCPartiallyMarkovs}, and the equality
\begin{align}
H(f(X,Y)|X,U,W) =0 .\label{eq:proofPartiallyFunction}
\end{align}

Then, based on $W$, we introduce a new random variable $W'$ such that  $U$ and $W'$ satisfy the inequalities of the theorem as well as the Markov chains \eqref{eq:rateCPartiallyMarkovs} and the relation \eqref{eq:rateCPartiallySets}, which completes the proof.

We start by showing that $U$ and $W$ satisfy the Markov chains
\begin{align}
&U-X-Y  \nonumber \\
X&-(U,Y)-W.\label{eq:proofPartiallyMarkovs}
\end{align}
For the first Markov chain, we have
$$H(Y|X,U)=\sum \limits_{q=1}^n \frac{1}{n} H(Y_q|X_q,U_q,q)\stackrel{(a)}{=}\sum \limits_{q=1}^n \frac{1}{n}H(Y_q|X_q)=H(Y|X),$$
where $(a)$ is due to the Markov chain $U_q-X_q-Y_q$. 

For the second Markov chain, we have
$$I(X;W|U,Y)\stackrel{(a)}{=}\sum \limits_{q=1}^n \frac{1}{n} I(X_q;W_q|U_q,Y_q)=0,$$
where $(a)$ is due to the Markov chain $X_q-(U_q,Y_q)-W_q$.

Equation \eqref{eq:proofPartiallyFunction} holds due to
\begin{align}
H(f(X,Y)|X,U,W)= \sum \limits_{q=1}^n \frac{1}{n}H(f(X_q,Y_q)|X_q,U_q,W_q)\leq \varepsilon_n \nonumber
\end{align}
and the fact that $\varepsilon_n$ can be chosen arbitrarily small. 

Finally to show that $U$ and $W$ satisfy the inequalities of the theorem, consider the following equalities
\begin{align*}
I(X;U|Y)&=\frac{1}{n}\sum \limits_{q=1}^n I(X_q;U_q|Y_q) \\
H(X|U,W)&=\frac{1}{n}\sum\limits_{q=1}^n  H(X_q|U_q,W_q) \\
I(Y;W|X,U)&=\frac{1}{n}\sum \limits_{q=1}^n I(Y_q;W_q|X_q,U_q) \\
H(X)+I(Y,W|X,U)&=\frac{1}{n}\sum \limits_{q=1}^n H(X_q)+I(Y_q,W_q|X_q,U_q).
\end{align*}
This, together with \eqref{eq:proofPartiallyIneq1}, \eqref{eq:proofPartiallyIneq2}, \eqref{eq:proofPartiallyIneq3}, and \eqref{eq:proofPartiallyIneq4} shows that $U$ and $W$ satisfy the inequalities of the theorem. 

Until here we have shown that $U$ and $W$ satisfy the inequalities of the theorem, the Markov chains \eqref{eq:rateCPartiallyMarkovs}, and the equation \eqref{eq:proofPartiallyFunction}. 

The last step consists in defining a new random variable $W'$ such that  $U$ and $W'$ satisfy the inequalities of the theorem, the Markov chains \eqref{eq:rateCPartiallyMarkovs}, and equality \eqref{eq:rateCPartiallySets}, which completes the proof. To do this we need the following definition.
\begin{definition} [Support Set of a Random Variable with Respect to Another Random Variable]\label{def:SupportSet}
Let $ (V,X) \sim p(v,x) $ where $V$ is a random variable taking values in some countable set $\mathcal{V}=\{v_1,v_2,\cdots\}$. The support set of $X$ with respect to $V$ is the random variable $S_X(V)$ defined as 
$$S_X(v_j)=(j,s=\{x : p(v_j,x)>0\})\quad v_j\in \mathcal{V}\,.$$
\end{definition}
 Note that $V$ and $S_X(V)$ are in one-to-one correspondence by definition. 
In the sequel, with a slight abuse of notation we write $Z \in S_X(V)$ whenever 
$Z\in S$ and write $S_X(V) \in \mathcal{A}$ whenever 
$S \in \mathcal{A}$	.

Let $W'=S_{(U,Y)}(W)$. According to Definition \ref{def:SupportSet} and relations \eqref{eq:proofPartiallyFunction} and \eqref{eq:proofPartiallyMarkovs}, $U$ and $W'$ satisfy
\begin{align}
X&-(U,Y)-W' \nonumber \\
H(f(&X,Y)|X,U,W') =0 \label{eq:proofPartially11}
\end{align}
and the inequalities of theorem. To conclude the proof it remains to show that
$$(U,Y) \in W' \in \text{M}(\Gamma(G_{U,Y|X,U})).$$
Note that $(U,Y) \in W'$ follows directly from the fact that $W'=S_{(U,Y)}(W)$. We show that $ W' \in \text{M}(\Gamma(G_{U,Y|X,U}))$ by contradiction. Suppose that $w' \in \mathcal{W}'$ is not an independent set in $G_{U,Y|X,U}$. Notice that for any $u_i,u_j \in \mathcal{U}$, with $u_i \neq u_j$, and $y_i,y_j\in \mathcal{Y}$, $(u_i,y_i)$ and $(y_j,y_j)$ are not connected in $G_{U,Y|X,U}$. Hence, there exists some $u \in U$ and $y_i,y_j \in \mathcal{Y}$ such that $(u,y_i),(u,y_j) \in w'$, \textit{i.e.},
\begin{align}
p(u,y_i,w')\cdot p(u,y_j,w')>0\,. \label{eq:proofPartially2}
\end{align}
Now, $(u,y_i)$ and $(u,y_j)$ are connected in $G_{U,Y|X,U}$. This means that there exists some $x \in \mathcal{X}$ such that
\begin{align}
p(x,&u,y_i)\cdot p(x,u,y_j)>0 \nonumber \\
&f(x,y_i) \neq f(x,y_j) \,.\label{eq:proofPartially3}
\end{align}
The relations \eqref{eq:proofPartially2}, \eqref{eq:proofPartially3} and the Markov chain $X-(U,Y)-W'$, imply that
\begin{align}
&p(x,u,w')>0 \nonumber \\
p(y_i|x,u&,w')\cdot p(y_j|x,u,w')>0 \nonumber \\
f&(x,y_i) \neq f(x,y_j)\,. \nonumber
\end{align}
From these relations one concludes that
$$H(f(X,Y)|X,U,W')\geq H(f(X,Y)|X=x,U=u,W'=w')\cdot p(x,u,w')>0,$$
which contradicts \eqref{eq:proofPartially11}.
\end{IEEEproof}
\begin{IEEEproof}[Proof of Theorem \ref{rate:R0} ]
For achievability, by setting $U=X$, $V=Constant$, $W=f(X,Y)$, and by using the Markov chain $T-X-f(X,Y)$ gives the desired result. Note that in this case the cardinality bound can be tightened by a standard use of Caratheodory's theorem.

 The converse can be found in the proof of \cite[Theorem~2]{AhlsKor75}.
\end{IEEEproof}
\begin{IEEEproof}[Proof of Theorem \ref{rate:RX} ]
From the converse of \cite[Theorem~3]{OrliRoc01}, we deduce that if a rate pair $(R_0,R_Y)$ is achievable, then there exist random variables $U$ and $W$ that satisfy \eqref{eq:twoRoundRates} and 
$$U-X-Y$$
$$X-(U,Y)-W$$
$$H(f(X,Y)|X,U,W)=0.$$

Finally, the same argument as the final argument of the converse proof of Theorem \ref{rate:partially} shows that $W'=S_{U,Y}(W)$ satisfies the above relations, the inequalities of the theorem, and 
$$(U,Y) \in W' \in  \Gamma (G_{U,Y|X,U}).$$

The cardinality bounds for $\mathcal{U}$ and $\mathcal{W}$ can be derived using the same method as the one used in the proof of Theorem~\ref{achiev} for bounding the cardinalities of $\mathcal{T}$ and $\mathcal{W}$, respectively.
\end{IEEEproof}

{\small{
\bibliographystyle{plain}
\bibliography{Bibliography}}}

\appendix[Proofs of Example \ref{ex:KaspiBerger2} claims 1., 2., and 3.]

In this Appendix we prove the claims stated in Example \ref{ex:KaspiBerger2}.

\begin{itemize}

\item[1. ] Suppose $T'$, $V'$ and $W$ satisfy the conditions of \cite[Theorem 5.1]{KaspBerg82}, \textit{i.e.},
\begin{align}
&T'-X-Y \nonumber \\
V'-(X&,T')-(T',Y)-W \label{eq:ex2Markovs1}
\end{align}
 and that there exist functions $g_1(T',V',W)$ and $g_2(T',V',W)$ such that
\begin{align}
\mathbb{E}d_1(X,g_1(V',T',W)) &=0 \nonumber \\
\mathbb{E}d_2(Y,g_2(V',T',W)) &=0. \label{eq:ex2Distortions1}
\end{align}

With this choice of auxiliary random  variables $T'$, $V'$, and $W$ the sum-rate constraint in \cite[Theorem 5.1]{KaspBerg82}  becomes
\begin{align}
R_X+R_Y > I(X,Y;V',T',W) \label{eq:ex2SumRate1}.
\end{align}
The minimum of the right-hand side of \eqref{eq:ex2SumRate1} over $(T',V',W')$ can be restricted to the case where $V$ is a constant. To see this, replace $T'$ by $T\defeq (T',V')$ and let $V$ be a constant. The random variables $T$, $V$, and $W$ satisfy \eqref{eq:ex2Markovs1} and \eqref{eq:ex2Distortions1} and give the same rate constraint as in \eqref{eq:ex2SumRate1}.

We now want to find the minimum of 
\begin{align}
I(X,Y;T,W)\label{eq:ex2Minimum}
\end{align}  for some $T$ and $W$ that satisfy
\begin{align}
&T-X-Y \nonumber \\
X&-(T,Y)-W \label{eq:ex2Markovs}
\end{align}
 and such that there exist functions $g_1(T,W)$ and $g_2(T,W)$ such that
\begin{align}
\mathbb{E}d_1(X,g_1(T,W)) =0 \nonumber \\
\mathbb{E}d_2(Y,g_2(T,W)) =0. \label{eq:ex2Distortions}
\end{align}
Using similar arguments as the one used for establishing the cardinalities in the proof of Theorem~\ref{achiev-rd1}, one can derive the following bounds for $\mathcal{T}$ and $\mathcal{W}$:
\begin{align*}
|\mathcal{T}|&\leq 4 \\
|\mathcal{W}|&\leq 9.
\end{align*}
The minimum of \eqref{eq:ex2Minimum} with the above cardinality bounds can in principle be numerically evaluated to obtain $\min I(X,Y;T,W)=1.03$. However, the number of degrees of freedom in the minimization still makes the problem intractable on a regular desktop computer. As it turns out, for the problem at hand  the cardinality bound $|\mathcal{W}|\leq 9$ can be tightened to $|\mathcal{W}|\leq 2$, which then allows to obtain the desired minimum in a matter of seconds. More details can be found in \cite{Sefi13}.
\item[2. ] The sum-rate constraint in \cite[Theorem 5.4]{KaspBerg82} is
\begin{align}
R_X+R_Y &> I(X,Y;T,W)\nonumber 
\end{align} 
for some $T$ and $W$  that satisfy
\begin{align*}
T-X-Y, 
\end{align*}
and such that there exist functions $g_1(W)$ and $g_2(W)$ such 
that
\begin{align}
\mathbb{E}[d_1(X,g_1(W))]&=0 \nonumber \\
\mathbb{E}[d_2(Y,g_2(W))]&=0. \label{eq:ex2FullCooperationDist}
\end{align}
Since $I(X,Y;T,W)\geq I(X,Y;W) $, by letting $T$ be a constant decreases the sum-rate constraint. We now want to find the infimum of $ I(X,Y;W)$ over 
$W$'s  that satisfy \eqref{eq:ex2FullCooperationDist} for some $g_1(W)$ and $g_2(W)$.

The distortion criteria \eqref{eq:ex2FullCooperationDist} imply that for any $w \in W$ with $p(w)>0$, we should have
\begin{align*}
P(W=w|X=-1)\cdot P(W=w|X=+1)&=0 \\
P(W=w|Y=-1)\cdot P(W=w|Y=+1)&=0.
\end{align*}
Because of the symmetry of $X$ and $Y$, $I(X,Y;W)$ is minimized for the random variable $W \in \{w_-,w_+\}$ with probability distribution
\begin{gather}
\raisetag{-10pt}
p(w_-|x,y) = \begin{cases}
1 \mbox{  if } x=-1 \mbox{ or } y=-1\\
\frac{1}{2}  \mbox{ if } (x,y)=(0,0)\\
0 \mbox{ otherwise}
\end{cases} \nonumber
\end{gather}

\begin{gather}
\raisetag{-10pt}
p(w_+|x,y) = \begin{cases}
1 \mbox{  if } x=+1 \mbox{ or } y=+1\\
\frac{1}{2}  \mbox{ if } (x,y)=(0,0)\\
0 \mbox{ otherwise.}
\end{cases} \nonumber
\end{gather}
This $W$ satisfies the Markov chain and distortion criteria constraints of  \cite[Theorem 5.4]{KaspBerg82} and gives
$$\inf (R_X+R_Y)=I(X,Y;W)=1-\frac{1}{7}=0.85$$

\item[3. ] Let $T,V$ be constants and $U \in \{u_-,u_+\}$  have the probability distribution
\begin{gather}
\raisetag{-10pt}
p(u_-|x) = \begin{cases}
1 \mbox{  if } x=-1\\
\frac{1}{2}  \mbox{ if } x=0\\
0 \mbox{ otherwise}
\end{cases} \nonumber
\end{gather}

\begin{gather}
\raisetag{-10pt}
p(u_+|x) = \begin{cases}
1 \mbox{  if } x=+1 \\
\frac{1}{2}  \mbox{ if } x=0 \\
0 \mbox{ otherwise.}
\end{cases} \nonumber
\end{gather}

Let $W \in \{w_-,w_+\}$ have the probability distribution
\begin{gather}
\raisetag{-10pt}
p(w_-|u,y) = \begin{cases}
1 \mbox{  if } (u,y) \in\{(u_-,-1),(u_-,0),(u_+,-1)\}\\
0 \mbox{ otherwise}
\end{cases} \nonumber
\end{gather}

\begin{gather}
\raisetag{-10pt}
p(w_+|u,y) = \begin{cases}
1 \mbox{  if } (u,y) \in\{(u_+,+1),(u_+,0),(u_-,+1)\},\\
0 \mbox{ otherwise.}
\end{cases} \nonumber
\end{gather}

Random variables $T$, $U$, $V$, and $W$ satisfy the Markov chains and distortion criteria of Theorem~\ref{achiev-rd1}. These random variables give the sum rate
$$R_X+R_Y>I(X,Y;W)=1-\frac{1}{7}=0.85$$
with
$$R_0>I(X;U|Y)=0.38.$$
\end{itemize}
\end{document}